\documentclass[traditabstract]{aa}
\usepackage{graphicx}
\usepackage{times}
\usepackage{natbib}
\newcommand{\enzo}{\it{\small ENZO}}

\begin{document}
\title{Forecasts for the detection of the magnetised cosmic web from cosmological simulations}
\author{F. Vazza\inst{1,2}, C. Ferrari\inst{3}, M. Br\"{u}ggen\inst{1}, A. Bonafede\inst{1}, C. Gheller\inst{4}, P. Wang\inst{5}}
\offprints{F.Vazza \\ \email{franco.vazza@hs.uni-hamburg.de}}

\institute{Hamburger Sternwarte, Gojenbergsweg 112, 21029 Hamburg, Germany
\and INAF/Istituto di Radioastronomia, via Gobetti 101, I-40129 Bologna
\and  Laboratoire Lagrange, UMR7293, UniversitŽ Nice Sophia-Antipolis, CNRS, Observatoire de la C™te dÕAzur, F-06300 Nice, France\and CSCS, Via Trevano 131, CH-6900 Lugano, Switzerland
\and  NVIDIA, Santa Clara, US}

\date{Received / Accepted}

\authorrunning{F. Vazza, C. Ferrari, M. Br\"{u}ggen, A. Bonafede, C. Gheller, P. Wang}
\titlerunning{The magnetised cosmic web in radio}

\abstract{The cosmic web  contains a large fraction of the total gas mass in the universe but is difficult to detect at most wavelengths.
Synchrotron emission from shock-accelerated electrons may offer the chance of imaging the cosmic web at radio wavelengths. In this 
work we use 3D cosmological {\enzo}-MHD simulations (combined with a post-processing renormalisation of the magnetic field to bracket for missing physical ingredients and 
resolution effects) 
 to produce models of the radio emission from the cosmic web. In post-processing we study the capabilities of 13 large radio surveys to detect this emission. We find that surveys by LOFAR, SKA1-LOW and MWA have a chance of detecting the cosmic web, provided that the magnetisation level of the tenuous medium in filaments is of the order of $\sim 1$\% of the thermal gas energy.}

\maketitle



\section{Introduction}
\label{sec:intro}

In cosmological structure formation, tiny matter perturbations grow through supersonic accretion of smooth and cold ($T \leq 10^4 \rm K$) matter. In the process, matter gets shock heated, thereby converting the infall kinetic energy into thermal energy \citep{1972A&A....20..189S}. Cosmological simulations have been used to measure the distribution of Mach numbers in cosmic shocks using a variety of methods. It has been found that weak ($M \sim 2-4$) merger shocks inside galaxy cluster are responsible for most of the conversion of the kinetic energy into gas heating, while stronger external accretion shocks ($M \sim 10-10^3$) are localized in the peripheral regions of large-scale structures  \citep[e.g.][]{ry03,pf06,sk08,va09shocks,va11comparison,2012MNRAS.423.2325A,2014arXiv1407.4117S}. All mass in the Warm-Hot-Intergalactic-Medium (WHIM) with temperatures in the range $\sim 10^5-10^7 \rm K$ has been processed by accretion shocks. Up to $\approx 90$ percent of the total baryon and dark matter mass in the Universe is expected to reside in this gas phase \citep[][]{1999ApJ...514....1C,2001ApJ...552..473D}, which is extremely difficult to be observed at most wavelengths. Indeed, only a few tentative detections of the WHIM associated with cosmic filaments have been claimed in X-ray \citep[e.g.][]{2003A&A...410..777F,2008A&A...482L..29W,2010ApJ...715..854N,2013ApJ...769...90N} or at microwaves using the Sunyaev Zeldovich effect \citep[][]{2013A&A...550A.134P}.
 A few possible detections of shocks emitting in radio have been also reported \citep[][]{2002NewA....7..249B,2007ApJ...659..267K,2010A&A...511L...5G,2013ApJ...779..189F}. 
It has been pointed out that the accretion shocks around cosmological filaments may be detected with radio telescopes provided these shocks are efficient enough in accelerating electrons to relativistic energies \citep[][]{2004ApJ...617..281K,2011JApA...32..577B,2012MNRAS.423.2325A,va15ska,2015arXiv150101023G}. 
A few authors have studied the emission from electrons accelerated at the intersections of galaxy clusters and cosmic filaments \citep[e.g.][]{2004ApJ...617..281K,2009MNRAS.393.1073B,sk11}, yet the detections were mostly related to intermittent mergers of  galaxy clusters, rather than to stationary accretion shocks.
\citet{2002NewA....7..249B} used the numerical method of \citet{mi01} to simulate a scenario in which a filament shock causes the observed radio emission in the cluster ZwCl 2341.1+0000. In order to explain the 
observed radio emission, $\sim 2$ percent of the shock ram energy has to go into relativistic electrons, and a magnetic energy of $\sim 1.5 ~\rm \mu G$ is required, probably indicating that the emission is related to more standard merger shocks \citep[][]{2009A&A...506.1083V,2010A&A...511L...5G}.\\

More recently, \citet{2012MNRAS.423.2325A} studied the  emission from {\it primary} electrons accelerated at cosmological shock waves and concluded that filaments must host fewer radio objects than galaxy clusters, with an average of a few radio objects brighter than $10^{32} \rm erg/(s \cdot Hz)$ at 1.4 GHz within $(500 \rm ~Mpc/h)^3$, and that most of this emission should come from the intersections with surrounding galaxy clusters.  They  predicted a flux density of radio emission from filaments  of the order of $0.12 ~\rm \mu Jy$ at redshift $z \sim 0.15$ and at a frequency of 150 MHz. \\

We recently found that the detection of the biggest filaments of the cosmic web might be within reach of the Square Kilometer Array if the magnetic field there reaches $\sim 10-100 ~\rm nG$ \citep[][]{va15ska}. While these first results were based on the simple assumption of a fixed magnetic field for a representative filament of our sample, in this work 
we use MHD simulations  with {\enzo} \citep[][]{enzo14} and study a much larger volume. 

\begin{table}
\label{tab:sim}
\caption{List of the simulations run for this project. First column: box length the simulated volume;  second column: number of grid cells in the initial conditions;  third column: spatial resolution; fourth column: DM mass resolution; fifth column: note on the adopted physics (MHD or Cosmic Rays (see Appendix).}
\centering \tabcolsep 5pt 
\begin{tabular}{c|c|c|c|c}
 $L_{\rm box}$ & $N_{\rm grid}$ & $\Delta x$ & $m_{\rm DM}$ & note \\\hline
 [$Mpc/h$]  &              & [$kpc/h$]  &  $[M_{\odot}/h]$ & \\  \hline
 50 & $2400^3$ & $20.8$ & $9.8 \cdot 10^{5} $ & MHD \\
 100 & $1200^3$ & $83.3$ & $6.2 \cdot 10^{7} $ & MHD \\
 200 & $1200^3$ & $166.6$ & $4.9 \cdot 10^{8} $ & MHD \\
 300 & $2048^3$ & $146.5$ & $3.4 \cdot 10^{8} $ & CR \\
\end{tabular}
\end{table}

\section{Methods}
\label{sec:methods}

\subsection{Numerical simulations with {\enzo}}

{\enzo} is a highly parallel code for cosmological simulations, which uses a particle-mesh N-body method (PM) to follow the dynamics of the DM and offers a variety of MHD-hydro solvers to compute the evolution of cosmic gas. 

\begin{figure*}
\begin{center}
\includegraphics[width=0.7\textwidth]{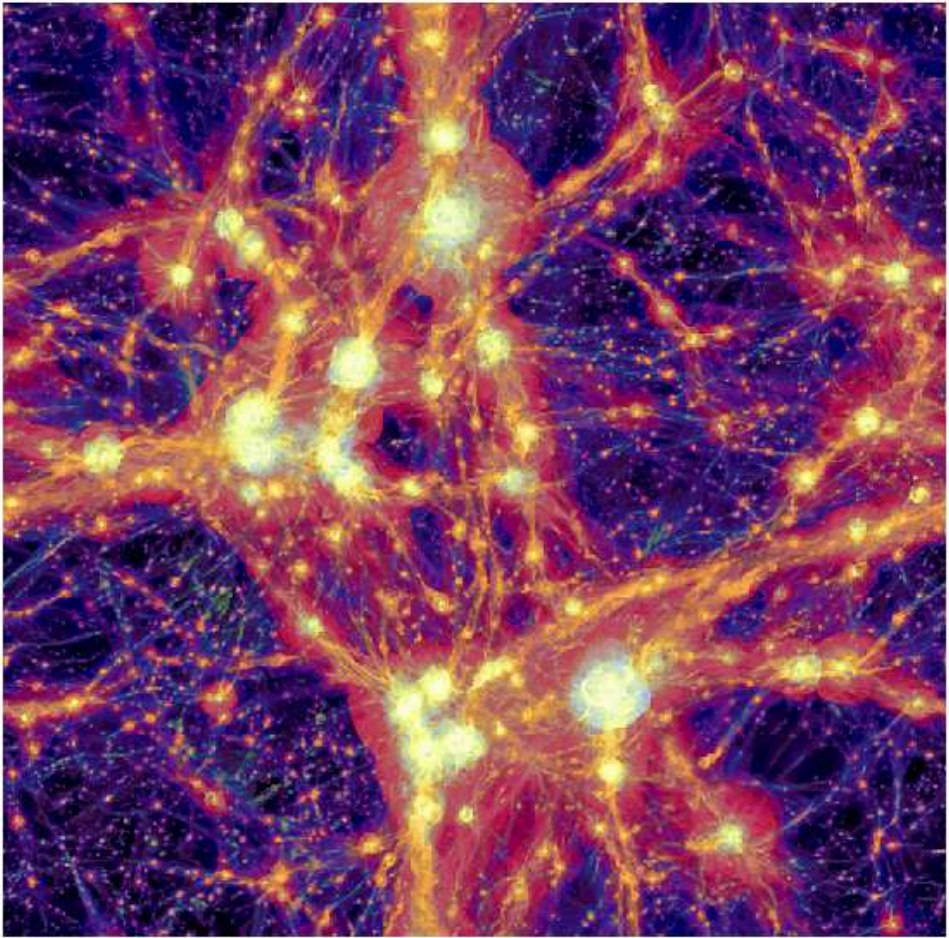}
\caption{3D rendering of the projected temperature (purple colours for $T \leq 10^{5} \rm K$ and red for $T \geq 10^{5} \rm K$) and magnetic field intensity (yellow for $B \geq 10 \rm nG$ and orange for $B < 10 \rm nG$) for our simulated $(50 \rm Mpc)^3$ volume at $z=0$, simulated with $2400^3$ cells and DM particles.}
\label{fig:3D}
\end{center}
\end{figure*}

Here we mostly analyze simulations obtained with ENZO-MHD described in \citet{va14mhd}, which belong to a large project run on Piz-Daint (Lugano, Switzerland) for a CHRONOS {\footnote {http://www.cscs.ch/user\_lab/allocation\_schemes/index.html}} proposal. 
This version of {\enzo} is based on the Dedner formulation of MHD equations \citep[][]{ded02}, which uses hyperbolic divergence cleaning to keep the $\nabla \cdot {\bf B}$ as small as possible during the computation. The MHD solver adopted here uses a piecewise-linear method reconstruction (PLM), where fluxes at cell interfaces are calculated using the local Lax-Friedrichs Riemann solver \citep[LLF][]{2000JCoPh.160..241K}. Time integration is performed using a total variation diminishing (TVD) second order Runge-Kutta (RK) scheme \citep[][]{1988JCoPh..77..439S}.  The PLM+MHD solver used in the simulations presented here (as well as the non-MHD version of the PPM hydro solver), has been ported to {\it Nvidia}'s CUDA framework, allowing {\enzo} to take advantage of modern graphics hardware \citep[][]{wang10,enzo14}.  In simulations using a single-level mesh, the GPUs provide a factor of $\sim 4$ in speed-up over CPU-only runs. Our run with the highest spatial resolution is a (50 Mpc)$^3$ volume simulated using 
$2400^3$ cells and $2400^3$ DM particles (resolution $20.8 ~ \rm kpc$), which, as far as we know, is the largest MHD cosmological simulation to date \citep[][]{va14mhd}.  This run used $\sim 4.5$ million core hours running on 512 nodes (2048 cores in total) on Piz Daint in Lugano.  Fig.~\ref{fig:3D} shows the projected (mass-weighted) temperature and magnetic field strength at $z=0$ for our best-resolved run ($2400^3$). The intricate network shows the outer shell of accretion shocks where kinetic energy is thermalised. In the inner regions, in the centres of halos and along the spine of filaments, the magnetic field gets amplified.
A list of the simulations analysed for this work is given in Tab.~1. 
To extend our results to a larger volume and include the role of secondary electrons injected by cosmic ray protons, we analyse a set of simulations that include CRs (details in the appendix and in \citealt{va14curie}).

\subsection{Cosmological Model}

We  assume a WMAP 7-year cosmology \citep[][]{2011ApJS..192...18K} with
$\Omega_0 = 1.0$, $\Omega_{\rm B} = 0.0455$, $\Omega_{\rm DM} =
0.2265$, $\Omega_{\Lambda} = 0.728$, Hubble parameter $h = 0.702$, a normalisation for the primordial density power
spectrum $\sigma_{8} = 0.81$ and a spectral index of $n_s=0.961$ for the primordial spectrum of initial matter
fluctuations. All runs start at $z_{\rm in} \geq 50$, the exact value of which varies with resolution.

\subsection{Non-thermal processes}

In order to model the synchrotron emission we have to know the energy spectrum of relativistic electrons and the magnetic field at each point in space and time. Both are largely unknown, specially at the low-density regime of cosmic filaments. In our analysis, we try to bracket the possible range by varying model parameters.


\subsubsection{Magnetic fields}
\label{subsubsec:Bfield}

The degree of magnetisation in the low-density cosmic web is largely uncertain.  The predictions from cosmological
simulations lie in the range of $\sim 10^{-4 } \rm \mu G-0.1 \mu$G for filaments, depending on numerical resolution of the cosmological simulation and as well as on the assumed seeding scenario \citep[e.g.][]{do99,br05,2003PhRvD..68d3002S,ry08, xu09, donn09,2015arXiv150600005M}. Since the growth of primordial magnetic fields in filaments is dominated by compressive motions and small-scale shocks \citep[][]{ry08,va14mhd}, the dynamical memory of the system might persist over long cosmological times and the observed magnetisation level closely connects to the seeding process(es). This is different from galaxy clusters where most of the magnetic energy is extracted from the kinetic energy, thereby quickly erasing previous dynamical information.
In our latest MHD runs we investigated the growth of magnetic fields starting from $B_0=10^{-10}$ G (comoving), seeded at high redshift  \citep[e.g.][and references therein]{wi11,ry11}. However, the  magnetic field measured even in our most massive clusters hardly reaches  $\sim 0.01-0.05  ~\mu G$, while in filaments the magnetic field is never larger than a few nG.
In clusters, the above values are $\sim 10-100$ times smaller than the typical magnetic field inferred using the Faraday Rotation effect \citep[e.g.][]{mu04,bo10,bo13}. Therefore, something appear to be missing in our numerical modeling of magnetic fields.
On the one hand, only within galaxy clusters we might have a large enough dynamical range to simulate a small-scale dynamo\citep[e.g.][]{2004ApJ...612..276S,2013NJPh...15b3017S}. In filaments, however,  the dominance of compressive forcing makes dynamo amplification inefficient. Even in high-resolution adaptive mesh refinement runs we did not measure any strong field growth beyond that caused by mere compression and magnetic flux conservation. 
On the other hand, additional seeding from astrophysical sources (star forming regions, galaxies and galaxy clusters) can magnetise large-scales, even if with unknown efficiency and duty cycles \citep[e.g.][]{Kronberg..1999ApJ,donn09,xu09,schober13,beck13}.  Fast-growing plasma instabilities, such as firehose and mirror instabilities, in large $\beta$ plasmas (where $\beta$ is the ratio between the thermal and the magnetic energy) might be able to amplify magnetic fields up to values we observe in clusters \citep[][]{sch05,2011MNRAS.410.2446K,2014MNRAS.440.3226M}, or at least to provide additional seeding for the dynamo. Indications of substantial magnetisation along the filamentary region in the SW sector of the Coma cluster have been reported in \citet[][]{bo13}.  Finally, it has also been suggested that cosmic-ray particles accelerated at strong shocks can amplify upstream magnetic fields substantially  \citep[][]{2012MNRAS.427.2308D, 2013MNRAS.tmp.2295B}. 

\begin{figure*}
\begin{center}
\includegraphics[width=0.45\textwidth]{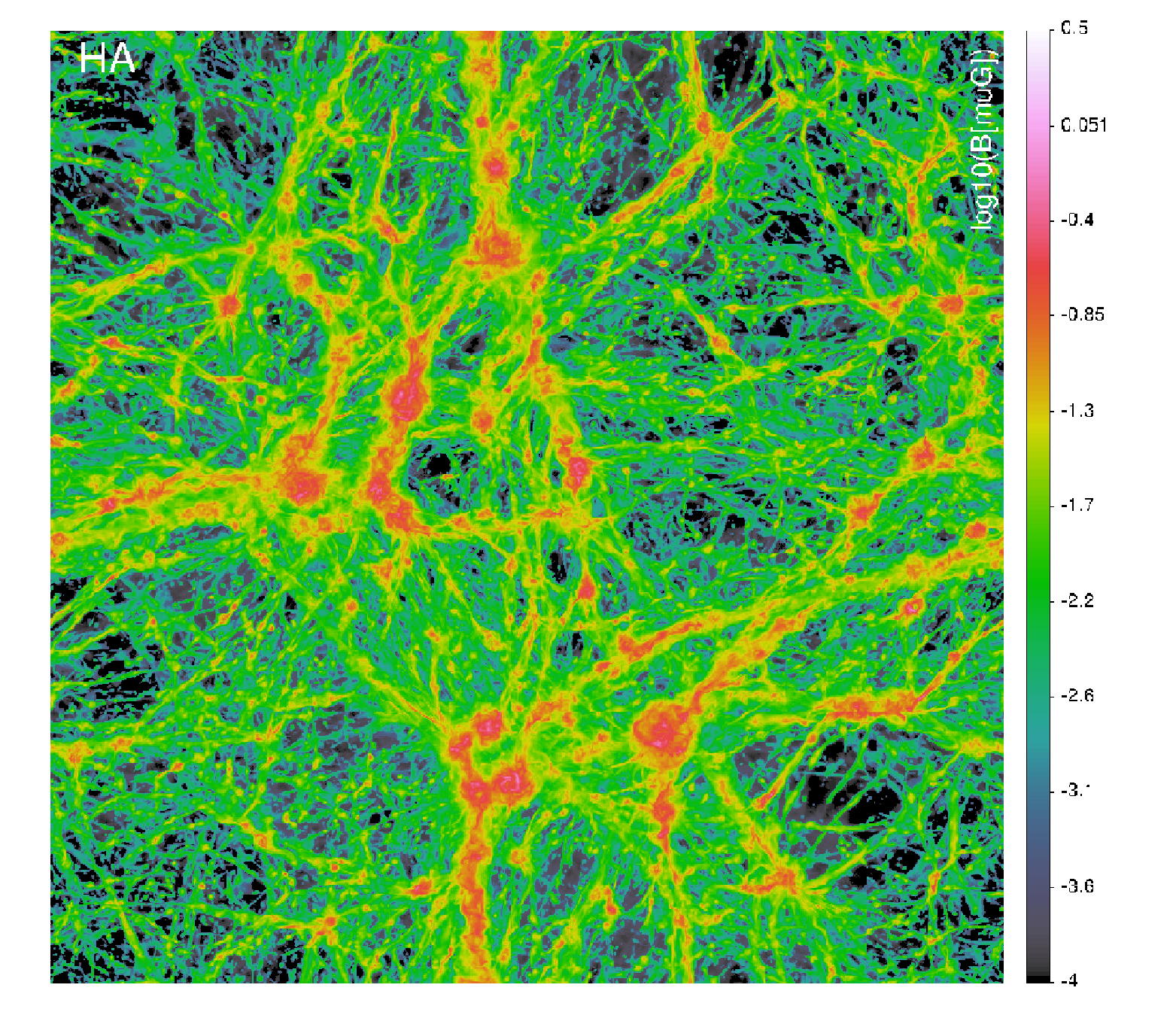}
\includegraphics[width=0.45\textwidth]{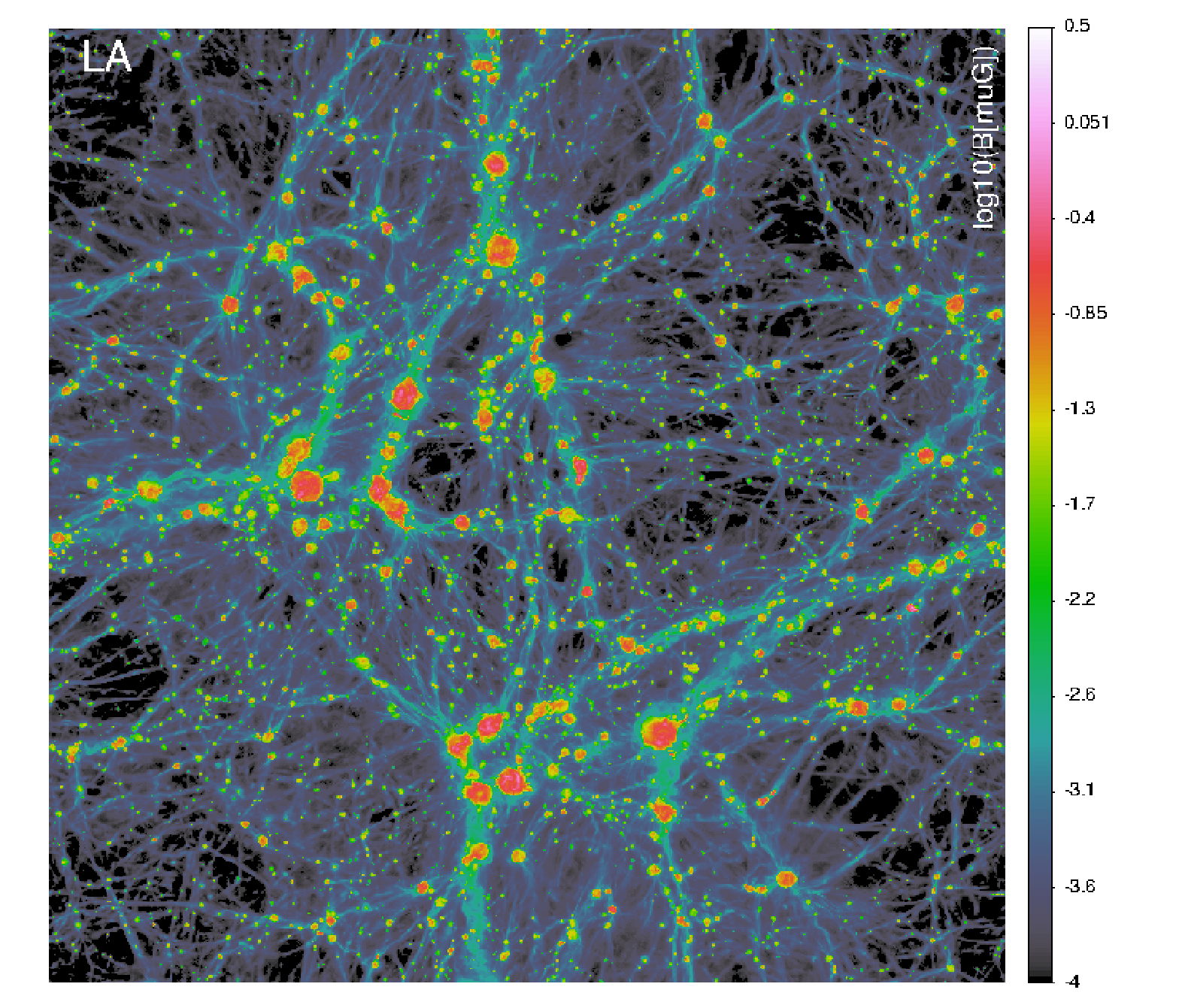}
\includegraphics[width=0.45\textwidth]{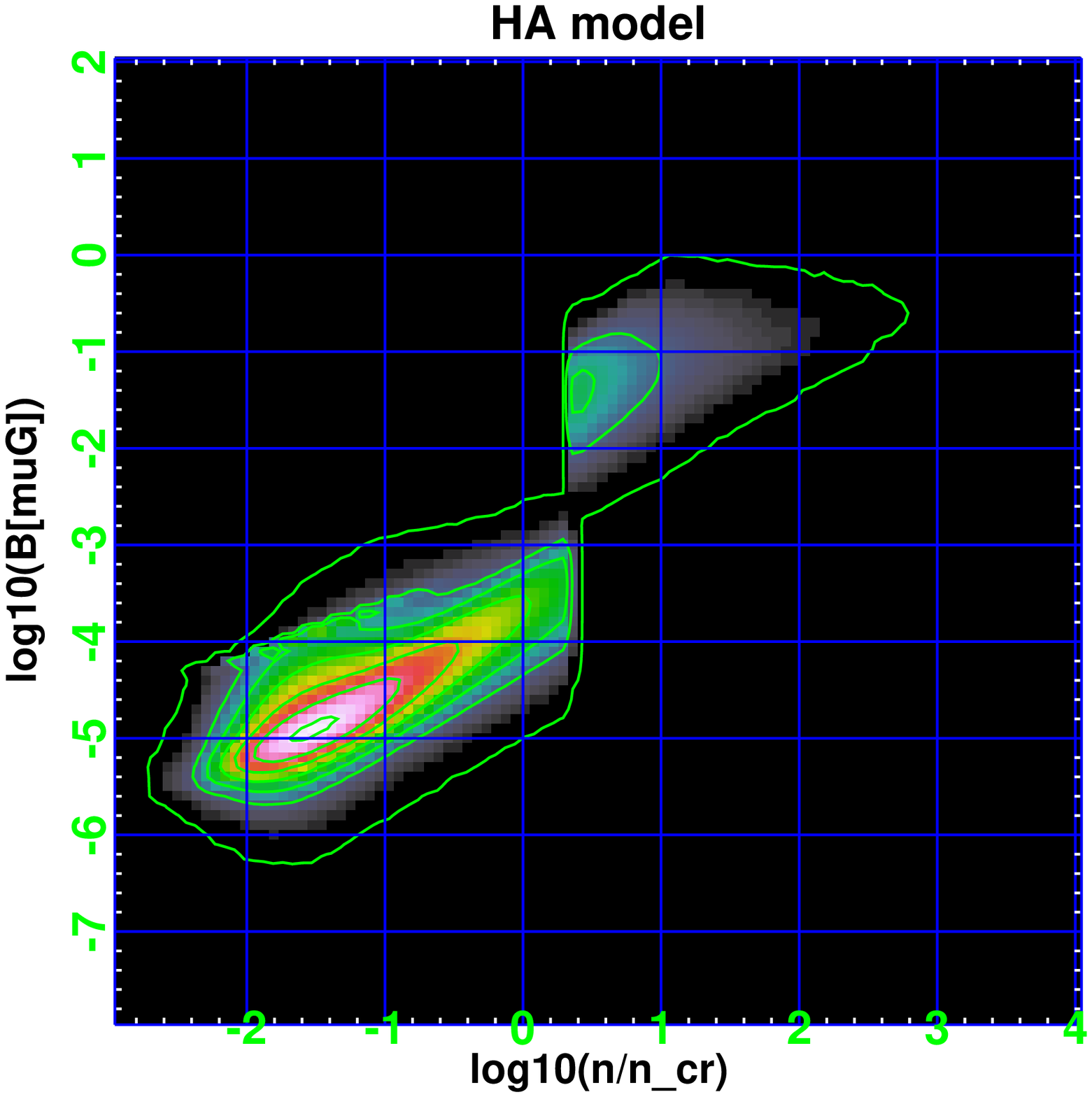}
\includegraphics[width=0.45\textwidth]{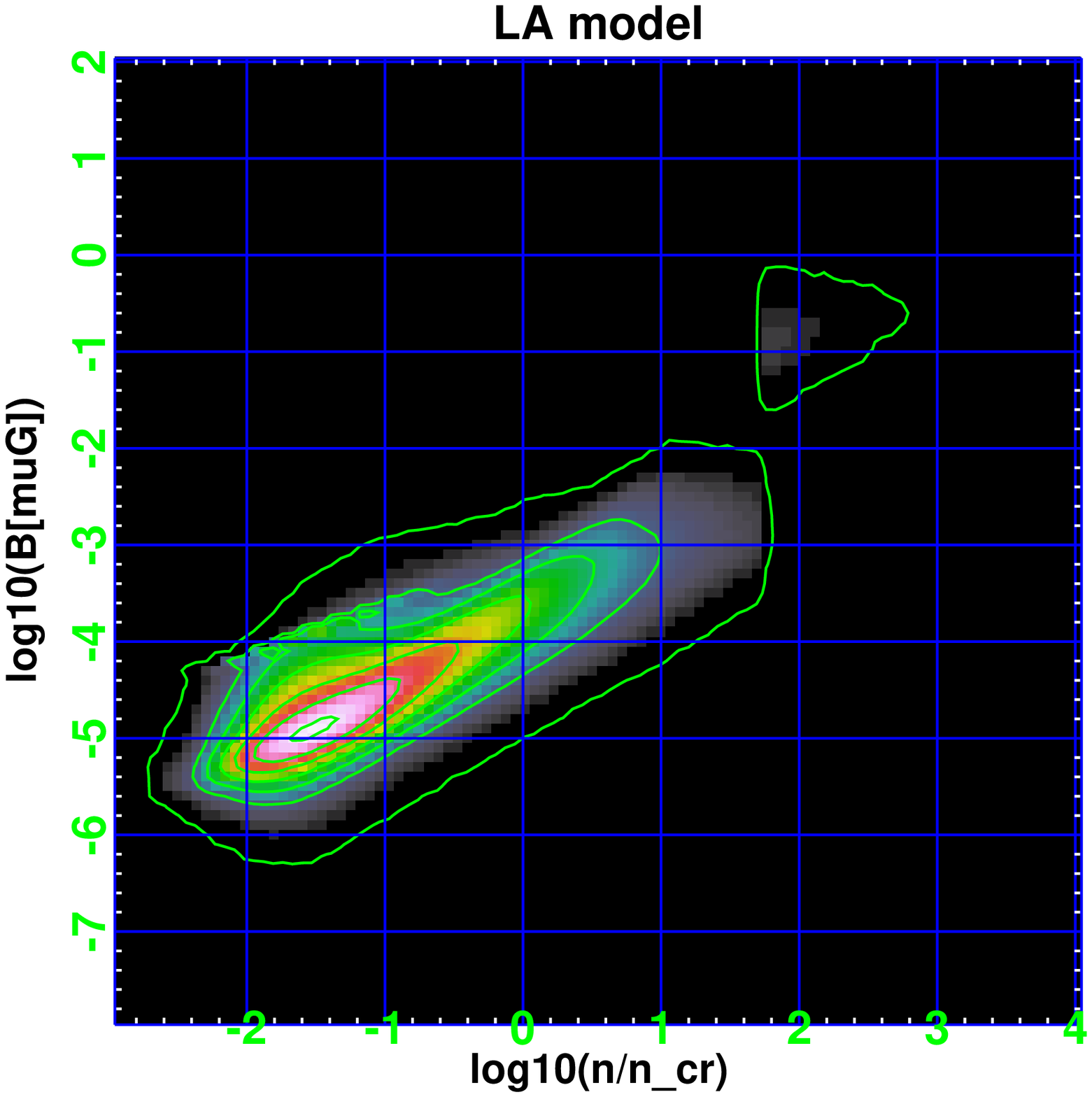}
\caption{Top: Mass-weighted magnetic field maps for the HA and LA model, for the same volume of Fig.~\ref{fig:3D}. The colour-bar is in units of $\log_{\rm 10}([\mu G])$. Bottom: phase diagrams on the  $(B,n/n_{\rm cr})$ plane for the same models. We additionally draw isocontours  tenfold  in the number of cells contributing to the pixels in the phase diagrams, starting from 100 cells.}
\label{fig:phaseB}
\end{center}
\end{figure*}

In summary, there is no firm conclusion about the strength of magnetic fields outside galaxy clusters, neither from observations nor from theory. Future radio observations will be crucial to assess this, and our goal here is to provide quantitative predictions for different magnetic field models, starting from extreme cases.
Hence, in post-processing we renormalize the magnetic field using two separate recipes: a high-amplification model (HA) and a low-amplification model (LA). In both cases, the normalisation depends on the local gas overdensity, $n/n_{\rm cr}$, where $n$ is the gas density and $n_{\rm cr,g}=3 H^2_{\rm 0}/(8 \pi G) \Omega_{\rm b}/\Omega_{\rm M}$ is the critical density rescaled for the gas fraction. 
In the HA case, we model the efficient magnetisation of all large-scale structures and increase the magnetic field for $n/n_{\rm cr}>2$. In the LA case, we limit the renormalisation to densities larger than the gas density at the virial radius of resolved halos in the simulation, $n/n_{\rm cr}>50$. In both cases, the strength of all components of the magnetic field in each cell is renormalized to $\beta= 100$, where $\beta$ is the ratio between the thermal energy and the magnetic energy. 
On the scale of clusters and filaments, the HA model typically yields a $\sim 10$ times larger magnetic field compared to what is predicted by our MHD runs, up to several $\sim \mu G$. The projected maps of magnetic fields and the $(B,n/n_{\rm cr})$ phase diagrams for these two models are given in Fig.~\ref{fig:phaseB}. In the density range of filaments ($n \sim 5-30 n_{\rm cr}$) the prediction of the two models differ significantly, i.e. $B \sim 0.001  ~\mu$G in the LA model (equivalent to our original MHD run for $n \leq 50 n_{\rm cr}$) and  $B \sim 0.05  ~\mu$G in the HA model. For $B \leq 3.2  ~\mu$G and flat electron spectra the synchrotron emission scales as $P_{\nu} \propto B^2$ (Eq.\ref{eq:hb} below) and, hence, this difference can have a large impact on the detectability of filaments in radio. Cells in the regime $n \sim 10-100 ~ n_{\rm cr}$ are more relevant for giant filaments and the outskirts of galaxy clusters, and the magnetic field varies from $0.001-0.01  ~ \rm \mu G$ (MHD model) to $0.1-1 ~ \rm \mu G$ (HA and LA). 
As we will show, the detectability of filaments in radio will crucially depend on what is the level of magnetisation at their typical overdensity, and therefore radio observation may have the potential of probing the mechanisms for magnetic field amplification in these regions.  
 
\begin{figure}
\begin{center}
\includegraphics[width=0.45\textwidth]{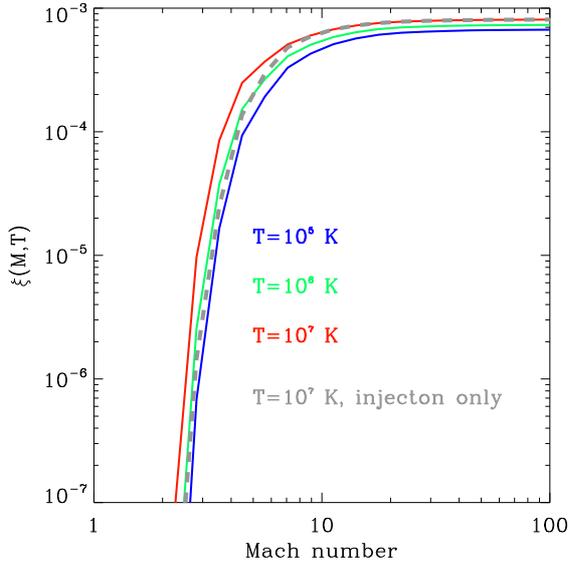}
\caption{The $\xi(M,T)$ function we use to compute the acceleration of CR-electrons at shocks, based on \citet[][]{hb07}.}
\label{fig:hb07}
\end{center}
\end{figure}

\subsubsection{The acceleration of relativistic particles at shocks}
\label{subsubsec:cr}

A plethora of mechanisms can contribute to the acceleration of electrons, e.g., direct injection from galactic winds and AGN, acceleration by cosmological shock waves, continuous injection by hadronic collisions and turbulent reacceleration. Simulations suggest that a vast majority of relativistic electrons in the WHIM should be produced by cosmological shock waves \citep[e.g.][]{ry03,pf06,2012MNRAS.423.2325A}, even though electron acceleration in high-$\beta$ plasmas is poorly understood \citep[e.g.][]{guo14}.
The theory of Diffusive Shock Acceleration (DSA) offers a theoretical framework to estimate the fraction of the kinetic energy
that shocks dissipate into the acceleration of  cosmic rays,  and links the slope of the produced momentum distribution to the shock Mach number \citep[e.g.][and references therein]{kr13}. However, the microphysical details are largely unknown and the efficiency of this process, specially at the weak shocks ($M \leq 5$) in clusters, are not understood \citep[e.g.][and discussion therein]{va14relics,va15relics}. In order to predict the synchrotron emission from accelerated electrons, we use the formalism by \citet[][]{hb07} that assumes an exponential cutoff in the energy distribution of electrons, determined by the balance of the acceleration rate from DSA and of the (synchrotron and Inverse Compton) cooling rate. In the downstream region, DSA is assumed to generate supra-thermal electrons that follow a power-law in energy.  The total emission in the downstream region is computed by summing up the contributions from all electrons accelerated at the shock. 

The monochromatic radio power at frequency $\nu$, $P_{\nu}$, depends on the shock surface area, $S$, the downstream electron density, $n_{\rm d}$, the electron acceleration efficiency, $\xi(M,T)${\footnote{We note that, in agreement with most of the literature on the DSA of electrons, we include in our definition of the electron acceleration efficiency the dependence on $M$ in Eq.\ref{eq:hb}, i.e. $\xi(M,T) = \xi_{\rm e,0} \Psi(M)$, where $\xi_{\rm e,0}=10^{-3}$ and $\Psi(M,T)$ is given in Eq. (31) of HB07.}} , the downstream electron temperature,  $T_{\rm d}$,  the spectral index of the radio emission, $s=\delta/2$, the magnetic field to the CMB energy density, $B_{\rm CMB}$ and the magnetic field in each cell, $B$:
\begin{equation}
P_{\nu}\propto S\, \cdot n_{\rm d} \cdot \xi(M,T) \cdot \nu^{-\delta/2} \cdot T_{\rm d}^{3/2} \frac{B^{1+\delta/2}}{(B_{\rm CMB}^2+B^2)} \ .
\label{eq:hb}
\end{equation}
Relic relativistic electrons resulting from previous shocks are expected to accumulate around $\gamma \sim 10^2$ \citep[][]{gb01} and can be re-accelerated by subsequent shocks at much higher efficiency. The net effect of freshly injected and re-accelerated particles can be modelled using the formalism by \citet[][]{kr13}, where the global effect of re-acceleration can be incorporated into the acceleration efficiency of CRs, to correct the $\xi(M,T)$ function of the HB07 model. It has been estimated in \citet{2013MNRAS.435.1061P} that for weak enough shocks in the ICM ($M \leq 3$) re-accelerated electrons dominate over directly injected electrons. For the sake of simplicity, we assume here that everywhere in the simulated volume the overall acceleration efficiency of CRs for $M \leq 3$ is affected by the re-acceleration of previously injected electrons. In this case the acceleration efficiency is  $\sim 10^{-5}-10^{-6}$ for $M=3$ (depending on the temperature) and 
$\sim 4-8 \cdot 10^{-4}$ for $M=10$.  Fig.~\ref{fig:hb07} gives the shape of the $\xi(M,T)$ function for different temperatures, and shows the role of shock re-acceleration for $M \leq 10$ shocks. Given the very steep dependence on Mach number of the acceleration efficiency at weak
shocks, the presence of re-accelerated particles can boost the net acceleration by one order of magnitude. 

Shock waves are identified in post-processing using the same procedure as in \citet{va09shocks}, which is based on the
analysis of velocity jumps of cells preliminary tagged based on the local 3D divergence. In the case of multiple shocked cells along the shock propagation axis, we perform a further cleaning in order to select only the cell yielding the
largest Mach number. To mask out some small differences in the reionization background, in post-processing we imposed a temperature floor of $T_{\rm re}=5 \cdot 10^3 \rm K$ everywhere \citep[][]{va09shocks}. The upstream/downstream shock parameters are then used in
Eq.~\ref{eq:hb} to compute the synchrotron emission as a function of frequency and build full-sky emission maps of our simulated volumes. The procedure is run in parallel for both magnetic field models of Sec.~\ref{subsubsec:Bfield} to produce a full catalog of mock radio observations. 

\begin{figure}
\begin{center}
\includegraphics[width=0.499\textwidth]{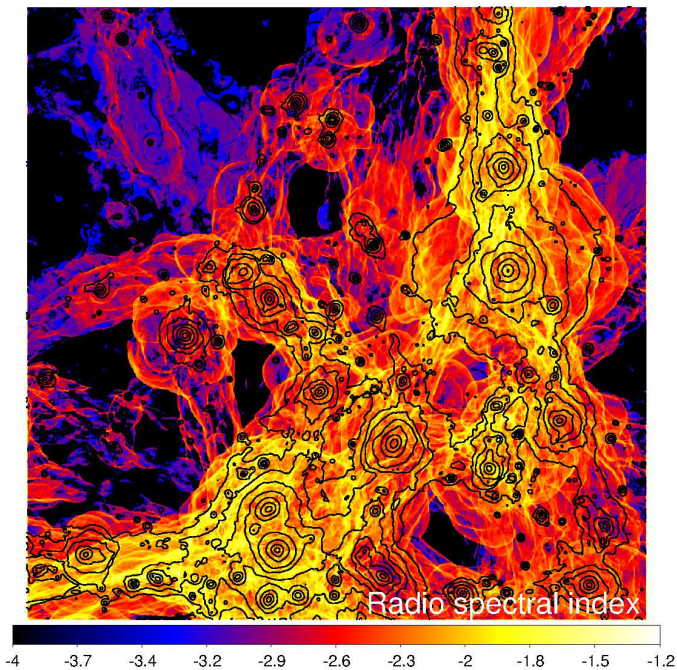}
\caption{Emission-weighted projected mean radio spectral index for a subregion with side $27 \rm ~Mpc$ (and depth $50 \rm ~Mpc$) through our highest resolution volume,
assuming the HA model for the magnetic field. The additional contours show the projected gas pressure for the same volume.}
\label{fig:spec}
\end{center}
\end{figure}

Fig.~\ref{fig:spec} shows the emission-weighted projected map of mean radio spectral index along the line-of-sight,  for our highest resolution box and assuming the HA model for the magnetic field. The accretion shocks have flat radio spectra, $s \sim 1-1.5$, while weaker shocks within clusters or filaments have steeper spectra, $s \approx 1.5-3$, as in the case of typical radio relics \citep[e.g.][]{hb07}.

\begin{table*}
\begin{center}
\caption{Fiducial parameters for our simulated radio observations. The quoted frequencies are reference values for each instrument. The last column gives the full resolution sensitivity we adopted in surface brightness, which is $3 \sigma$ of the assume thermal noise per beam (in the case of SKA1-LOW, this is the confusion noise). See Sec.~\ref{subsubsec:arrays} for details.} 
\centering
\begin{tabular}{|c|c|c|c|c|c|c|}

 \hline 
    array & configuration/strategy &  frequency & resolution  & min. baseline & sensitivity & detection threshold \\
          &  & $[\rm MHz]$ & $[\rm arcsec]$  & $[\rm m]$ & $\rm [mJy/beam]$ & $\rm [\mu Jy/arcsec^2]$\\
\hline
    VLA & VLSSr            & 74     & 80       & 35   & 100.0    & 42.365\\
    VLA & NVSS             & 1400   & 45        & 35   & 0.45  & 0.588\\
    Westerboork & WENSS    & 330    & 54        & 36   & 3.6   &  3.268\\
    GMRT & TGSS            & 150    & 20         & 100  & 5.0     &  33.098\\
    Molonglo & SUMSS       & 840    & 43       & 15   & 1.0    &  1.307\\
    LOFAR-HBA & Large Survey 1    & 120    & 25       & 40   & 0.25 & 1.059\\
    LOFAR-LBA & Large Survey     & 40     & 25        & 40   & 2.0 &  8.473\\
    LOFAR-HBA & Large Survey 2     & 120    & 5        & 40   & 0.1 &  10.591\\
    MWA & Broadband Survey & 150    & 120      & 7.7  & 10   & 1.838\\
    SKA1-LOW & Cont. Survey & 120 & 10  & 45 & 0.02 &  0.17 \\
    SKA1-MID & Band2 Wide Survey & 1000 & 0.5   & 15 & 0.001 & 10.591 \\
    SKA1-MID & Band2 Deep Survey & 1000 & 0.5   & 15 & 0.0002 & 2.118\\  
    ASKAP & EMU  & 1400 & 10  & 12 & 0.01 & 0.264 \\
 \hline 
\end{tabular}
\end{center}
\label{tab:surveys}
\end{table*}

\subsubsection{Mock radio observations}
\label{subsubsec:arrays}

In \citet[][]{va15ska} we explored the possibilities of future surveys and deep pointings with the SKA to detect the radio emission from filaments in targeted observations. In general, predicting the emission from a specific filaments is not trivial and the spatial distribution of gas in filaments in the surroundings of galaxy clusters is not known either.  In this work, we instead explore the potential of current and future radio surveys  to detect the diffuse radio emission from the cosmic web in a statistical way and without a priori knowledge of the best fields to target. Table 2 gives the full set of specification used to produce a mock observation of a simulated field, according to a procedure similar to \citet[][]{va15ska}:

\begin{enumerate}
\item We compute the radio emission at different redshifts produced during the simulation, in their comoving frame, and compute the total emission map by summing up the contribution of all cells along the line of sight. 
\item We convert the emitted power in the physical frame, assuming an observer at different redshift, i.e. we include the effect of the luminosity distance of the emitting frame and the effect of cosmological dimming. 
\item We transform the images in the Fourier space using FFT. We remove from the Fourier plane the frequencies smaller than the minimum antenna baseline of each specific radio configuration, mimicking the loss of signal from scales larger than the minimum instrumental baseline. This is particularly relevant for the large-scale diffuse emission we are investigating here.
\item After the baseline removal the images are converted back into real space. Then the emission is convolved for the beam using a 2D Gaussian filter. 
\item Only the pixels whose emission  is $I(x,y,z) \geq 3 \sigma$ are considered as detected by the specific observation. Depending on the array configuration, $\sigma$ is either the confusion limit (e.g. the contribution of unresolved background of continuum radio sources below the beam size) or the thermal noise of the instrument. The equivalent noise per beam is computed as $I_{\rm thr} \approx \sigma/(1.1333 W^2_{\rm Beam})$, where $W_{\rm Beam}$ is the beam size.
\end{enumerate}

In all cases, we implicitly assume a perfect removal of the Milky Way foreground \citep[e.g.][for the case of SKA observations]{2015MNRAS.447.1973B}, of all resolved point-like radio sources as well as an ideal calibration and deconvolution of the radio data. 
We post-process our simulated boxes under many different observing configurations, exploring the capabilities of existing and future surveys in radio. 
The past and current surveys are: a) VLSS Redux survey at $74$ MHz with the VLA \citep[][]{2012RaSc...47.0K04L}; b) NVSS survey at $1400$ MHz with the VLA \citep[][]{1998AJ....115.1693C}; c) the WENSS survey at $330$ MHz with Westerbork \citep[][]{1997A&AS..124..259R}; d) the TGSS survey at $150$ MHz with GMRT \citep[][]{2011ApJ...736L...8B}; e) the SUMSS survey at $840$ MHz with Molonglo \citep[][]{1999AJ....117.1578B}. 
For the future surveys it is more difficult to quote specific numbers. Thus, in some cases we explored a few possible configurations bracketing their specifications.
In detail, we studied: a) Large Area surveys with LOFAR HBA ($120$ MHz) and LBA ($40$ MHz) \citep[][]{2011JApA...32..557R}. The beam of LOFAR is formed digitally and the  final resolution/sensitivity that each survey will be able to reach is not yet fixed. For this reason, we studies two cases, based on \citet[][]{2014arXiv1412.5940C}; b) SKA1-LOW ($120$ MHz) and SKA1-MID ($1000$ MHz) surveys. Also in this latter case, we explored two possible SKA1-MID configurations, as in \citet[][]{2014arXiv1412.6512P}; c) Broad Band Survey with MWA at $150$ MHz, as in \citet[][]{2013PASA...30....7T}; d) EMU  survey with ASKAP at $1400$ MHz \citep[][]{2013PASA...30...20N}.

All used parameters are given in Tab.~\ref{tab:surveys}. The last column gives the detection level in surface brightness for each survey. In the case of SKA1-LOW, this is computed considering confusion noise. We also list the field-of-view of each instrument but this never entered our calculations since we consider full-sky surveys or very large surveys obtained by adding many pointings. 

A public repository of radio maps for the full volumes studied in this work is available at http://www.hs.uni-hamburg.de/DE/Ins/Per/Vazza/projects/Public\_data.html .

\begin{figure}
\begin{center}
\includegraphics[width=0.495\textwidth]{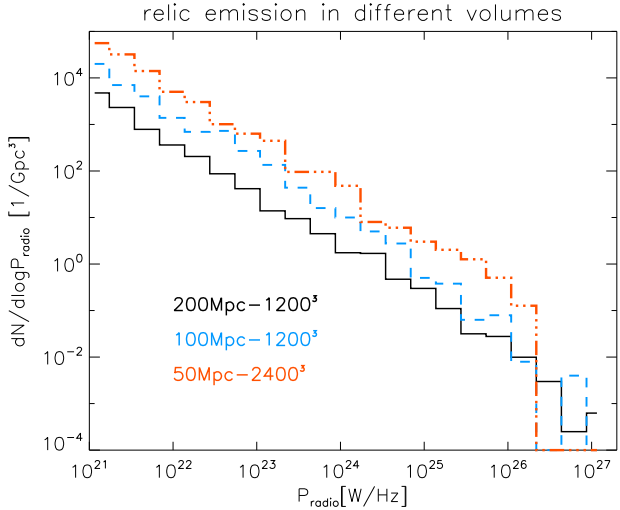}
\caption{Radio distribution function for the clusters in our three simulated volumes at $1.4$ GHz, using the LA model. No observational effects are considered in this case. }
\label{fig:resolution}
\end{center}
\end{figure}

\begin{figure}
\begin{center}
\includegraphics[width=0.495\textwidth]{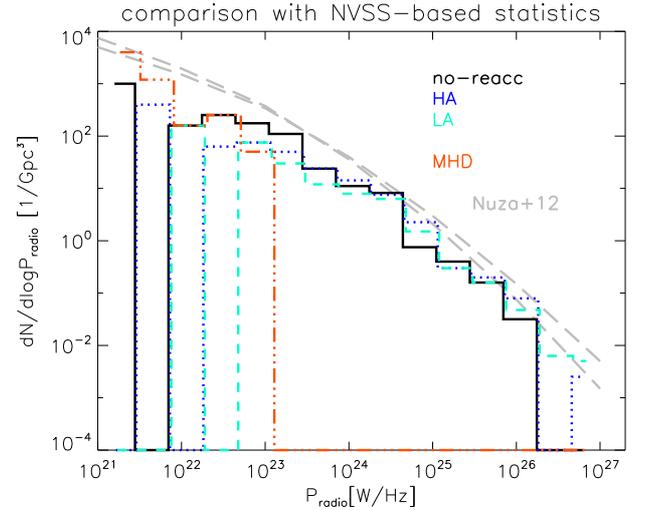}
\caption{Radio distribution function for the clusters in our $(100 ~\rm Mpc)^3$volume, normalized to $\rm Gpc^3$, for all investigated emission models. The additional gray lines show the range of the best fit to the distribution of radio relics based on the  NVSS, computed by \citet{2012MNRAS.420.2006N}.}
\label{fig:nvss}
\end{center}
\end{figure}

\begin{figure}
\begin{center}
\includegraphics[width=0.495\textwidth]{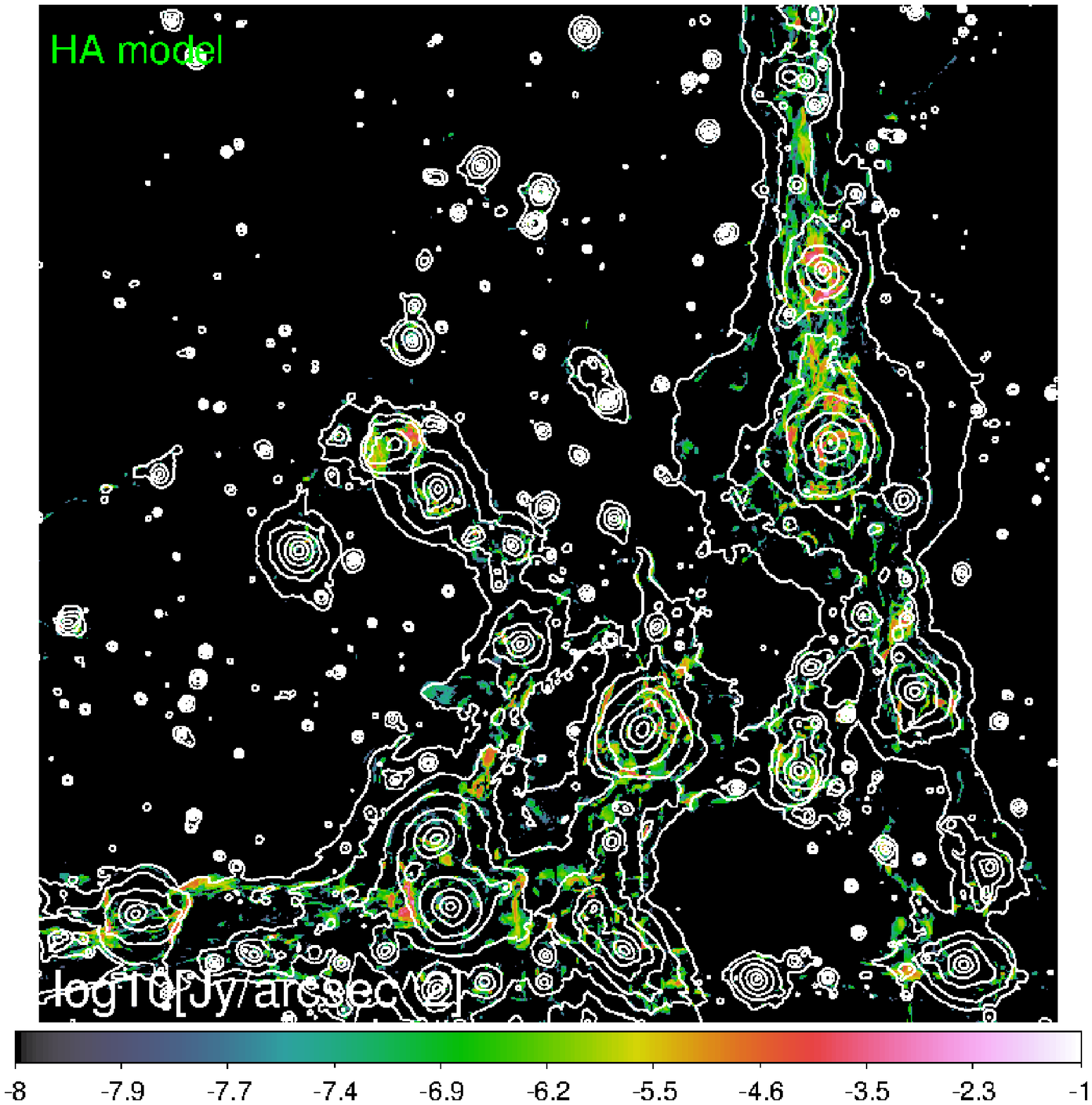}
\includegraphics[width=0.495\textwidth]{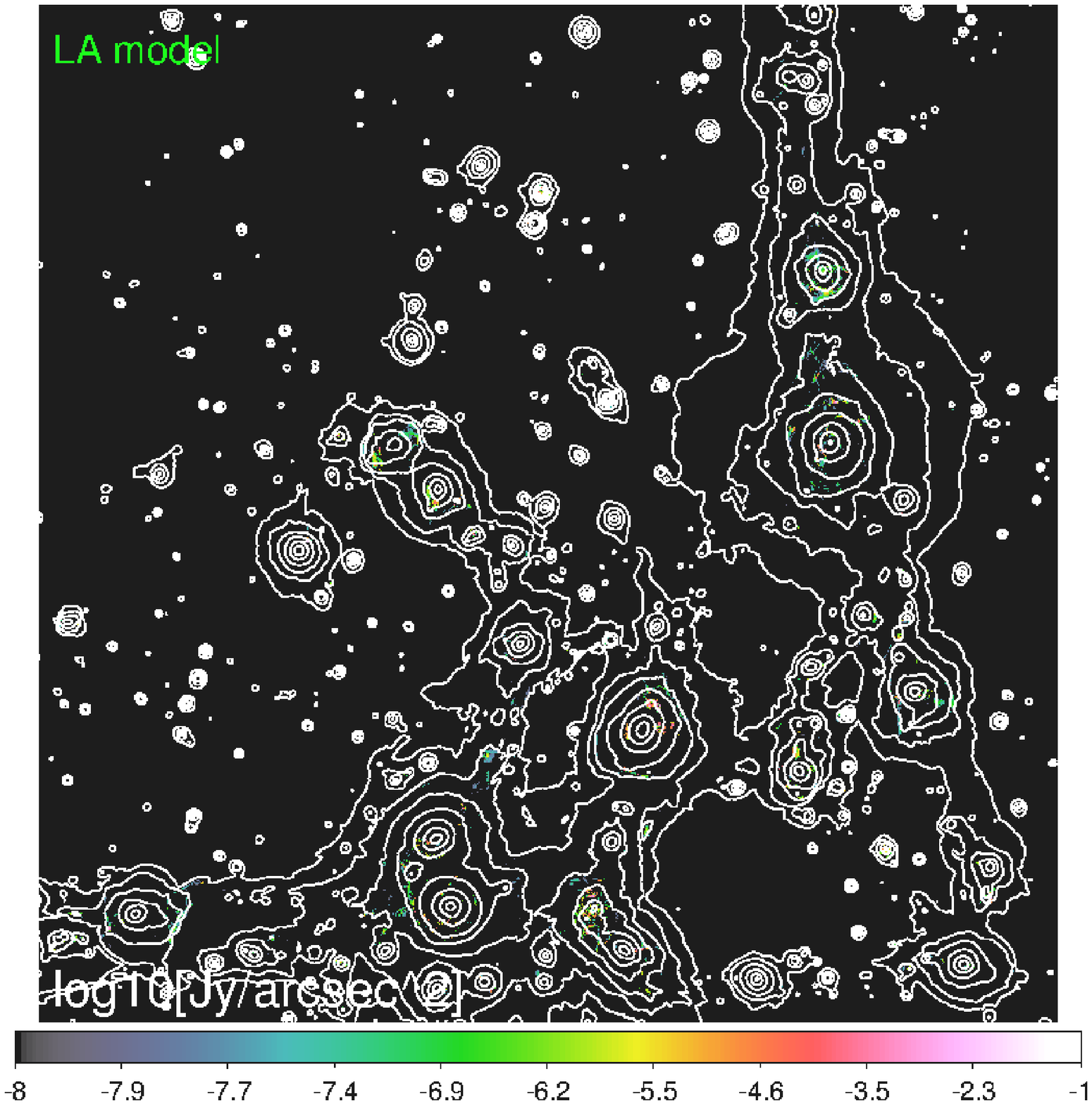}
\caption{Radio emission from a subregion with side $27 \rm ~Mpc$ (and depth $50 \rm ~Mpc$) through our highest resolution volume for the HA and LA magnetic fields models (units of $\log_{\rm 10}[$Jy/arcsec$^2]$ assuming a luminosity distance of 200 Mpc). The white contours show the projected gas pressure.}
\label{fig:radio_models}
\end{center}
\end{figure}

\section{Results}
\label{sec:res}

\subsection{Comparison with radio-relic emission in the NVSS}
\label{nvss}
First, we check our emission model against the
available constraints from the NVSS, using relics in galaxy clusters. The model by \citet{hb07} has been derived to model radio relics and in our case it fits rather well the 
distribution of relic emission in clusters. 
We compute the radio emission within $R_{\rm 200}$ ($\approx 0.7 ~R_{\rm vir}$) of  all identified clusters in the three simulated volumes at $z=0.05$ and compare the observed statistics of radio relic emission for the same reference magnetic field model (here the LA model), first without including observational effects in order to better focus on the effects of resolution and volume in our runs. 
The result is shown in Fig.\ref{fig:resolution}, where we renormalise the measured statistics to the same volume of $(1 ~\rm Gpc)^3$. Two clear trends stand out. First, the distribution functions of the different volumes have the same shape, with a factor of $\sim 10$ different normalisation in the number of objects when the high resolution and low resolution are compared. This follows from the fact that the internal dynamics (leading to merger shocks) of clusters is  better resolved at  higher resolution  \citep[][]{va11comparison}, producing more radio emitting objects. Second, the increase in the simulated volume enables the formation of more massive clusters, enabling more powerful major mergers. For this reason, our largest volume has a 
 $\sim 10$ times more powerful tail of high radio power.  The above trends are in line with what already known in the literature and based on shock statistics \citep[e.g.][]{ry03,pf06,va09shocks,va11comparison} and on simulated radio relics \citep[][]{sk11}. Having a simulation that combines the exquisite uniform resolution of our $2400^3$ run with the large volume of the $(200 Mpc)^3$ is presently unfeasible. In the following, we will explore the simulations we have presented here to extract as much as possible about cosmic filaments in radio emission.  We notice, however, that  weak merger shocks leading to relics are more subject to resolution effects in grid methods than strong accretion shocks, and that a resolution of $\sim 200  ~\rm kpc$
is already sufficient to capture the bulk of energy dissipation in large-scale shocks \citep[e.g.][]{ry03,va09shocks,va11comparison}. All runs presented in this work have a spatial resolution better than this, and the estimated radio power of our filaments is therefore a more robust statistics.

Second,  to better compare with the observed distribution of radio relics, we compute the radio distribution function for each magnetic field model in our intermediate $(100 ~\rm Mpc)^3$ volume at $z=0.05$, and compare it with the relic distribution
function based on the NVSS and derived by \citet{2012MNRAS.420.2006N}. This is so far the best guess for the abundance of radio relics in the local Universe, obtained by extrapolating the distribution function of radio relics in simulations, with adjustable parameters calibrated using all radio relics in the NVSS and with X-ray scaling relations.
Now we also introduce the observational effect discussed in Sec~\ref{subsubsec:arrays} and mimic the sensitivity of the NVSS survey.The result is shown in Fig.~\ref{fig:nvss}. As expected, the baseline MHD model cannot reproduce the distribution of bright radio relics, owing to its very low magnetic fields in clusters. 
The other models instead show a reasonable agreement with the statistics derived from NVSS, with significant gaps only at high radio power ($\geq 10^{26} \rm W/Hz$) due to the lack of massive objects, and and very low radio power ($P_{\rm radio} \leq 10^{23} \rm W/Hz$) due to the combined effect of limited spatial resolution to properly resolve the internal structure of the smallest clusters in the box, and of the loss of low-surface brightness objects when the observational cuts are introduced.
The differences between magnetic field models with post-processing amplification are very small because their differences only come into effect beyond the virial radii of clusters. 

Even if the DSA model for the  weak shocks responsible for relics is still under debate, it is reassuring that our models do not overestimate the counts of radio relics in the NVSS. We assume that the extrapolation of these models to the stronger accretion shocks around filaments and the typical condition of the WHIM is valid.

\subsection{Which level of magnetisation can be detected?}
\label{magnetic}

We start by focusing on the minimum magnetic field strength that can enable a detection of the cosmic web at any redshift.
Fig.~\ref{fig:radio_models} gives the total radio emission found with the HA and LA model for our most resolved run ($50^3 ~\rm Mpc^3$) with a resolution of $20.8$ kpc), with additional contours of gas pressure to trace the distribution of baryons. In both models the strongest emission patterns trace merger shocks inside clusters, producing structures similar to giant radio relics. The flux densities from these relics reaches $\sim 0.01-0.1$ Jy/arcsec$^2$ in the HA model, $\sim 10^{-5}-10^{-4}$ Jy/arcsec$^2$ in the LA model.
At the accretion shocks around clusters and filaments the differences are similar: the flux densities typically reach $\sim 10^{-8}$ Jy/arcsec$^2$ in the HA model and $\sim 10^{-11}$ Jy/arcsec$^2$ in the LA model. No significant differences are found at the scale of filaments when the re-acceleration of electrons at shocks is neglected (not shown), which means that filaments only produce radio emission where $M \geq 10$.\\

\begin{figure*}
\begin{center}
\includegraphics[width=0.85\textwidth]{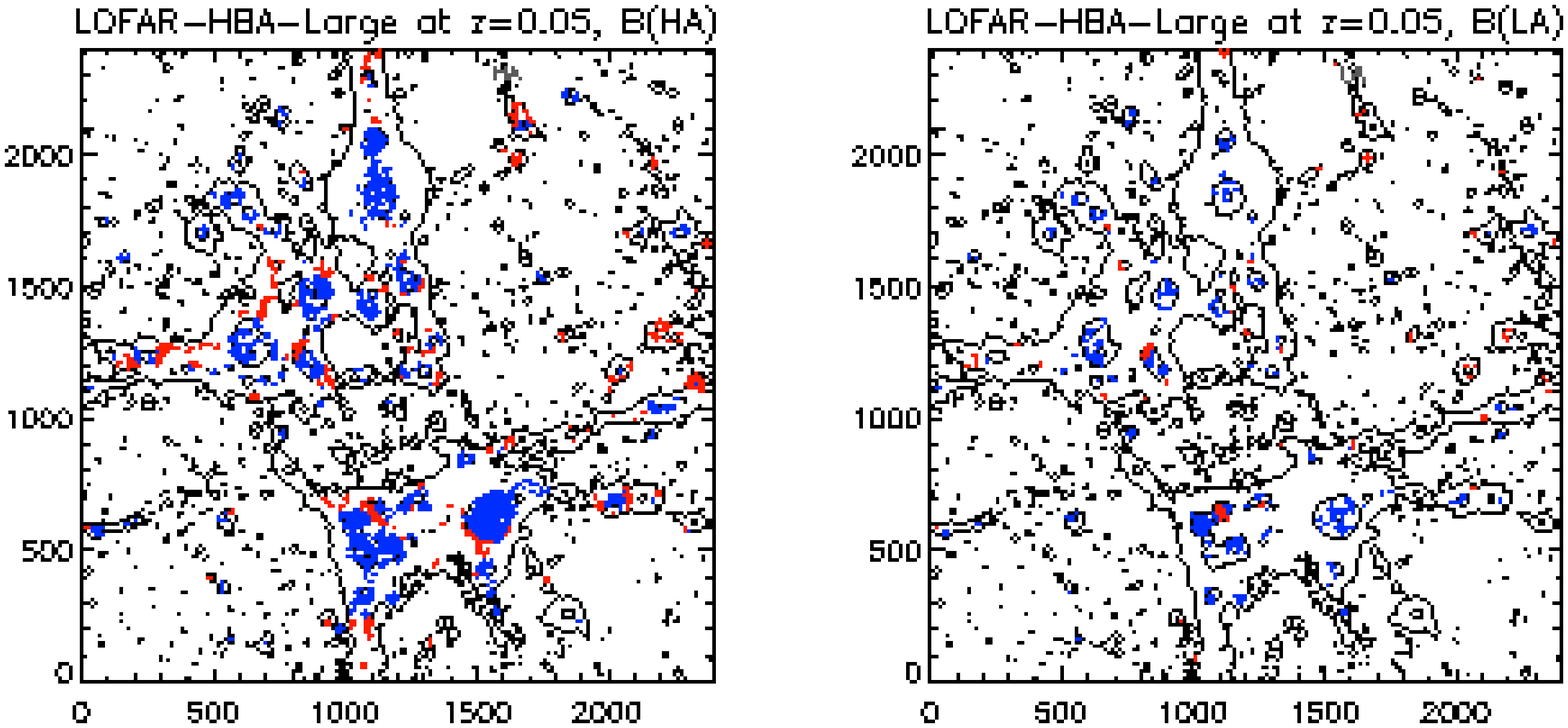}
\includegraphics[width=0.85\textwidth]{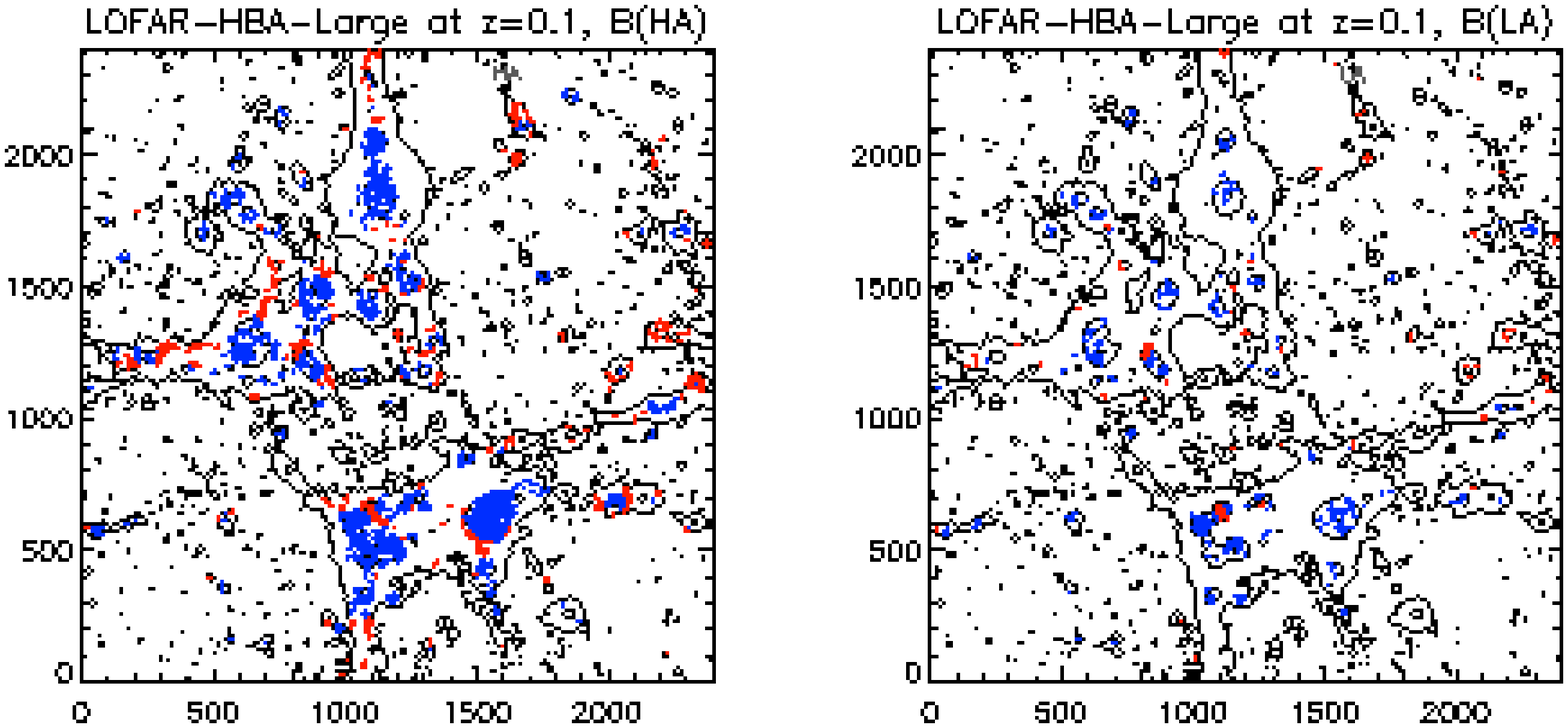}
\includegraphics[width=0.85\textwidth]{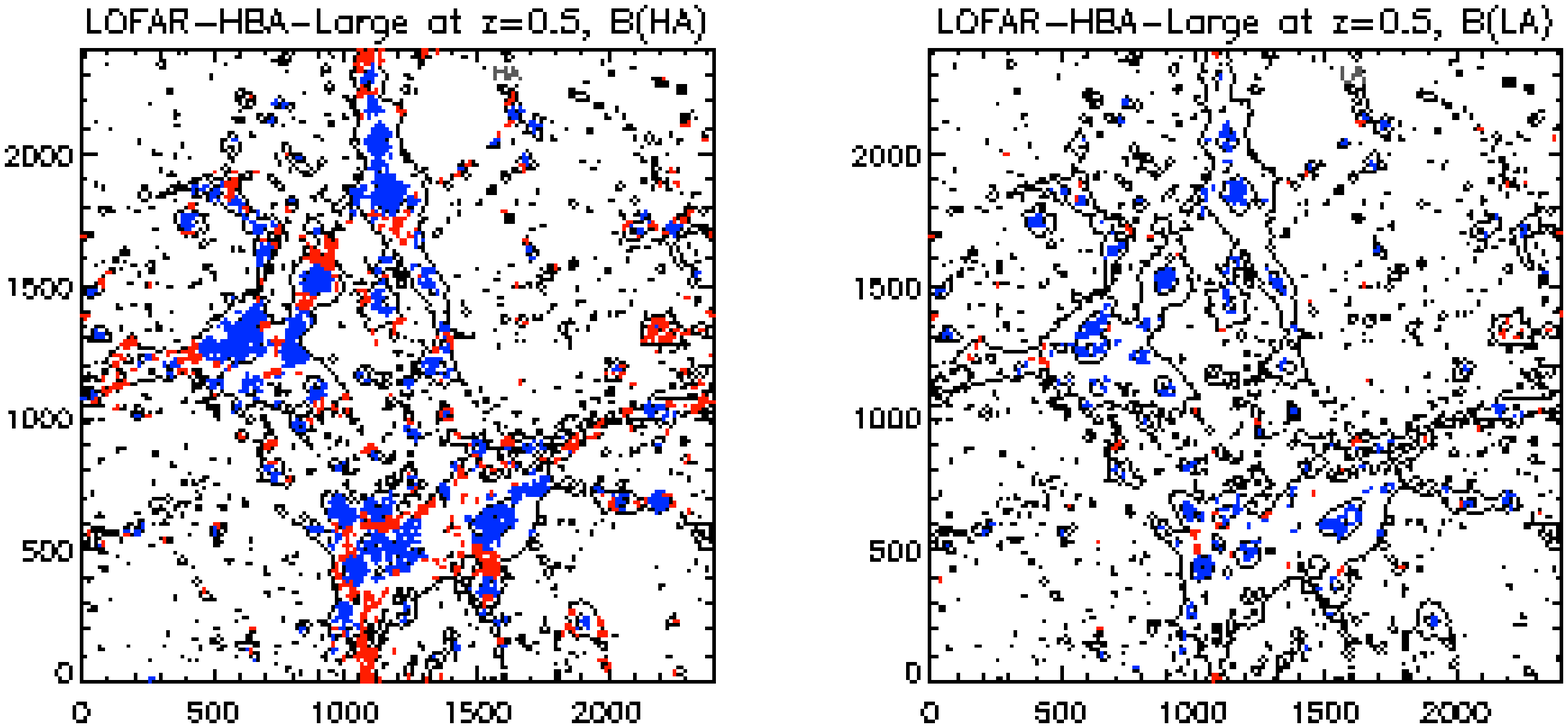}
\caption{Mock observation of the (50 Mpc) $^3$ volume using the configuration of LOFAR-HBA (Large Survey1, see Tab.~\ref{tab:surveys}) for increasing redshift ($z=0.05$, $z=0.1$ and $z=0.5$) for the HA (right) and LA (left)  magnetic field models.   The blue and red regions show the emission that is detectable for each model/redshift, divided into emission coming from $\leq 1.5 \rm r_{\rm vir}$ (blue) of each identified halo, or generated outside and therefore likely related to filaments (red). The additional contours show the projected gas pressure. The units of the axis are given in cell units ($\Delta x= 20.8 \rm ~kpc$).}
\label{fig:sur1}
\end{center}
\end{figure*}

Next we compute how much of the above emission patterns can be detected by real radio telescopes, and we start with one of the most promising cases explored here,  the Large Survey with LOFAR-HBA  (first configuration in Tab.~2, assuming a noise per beam of 0.25 mJy and a resolution beam of $25"$) at $120$ MHz  (Fig.~\ref{fig:sur1}). The region covers a projected area of $(50  ~\rm Mpc)^2$  with $\sim (205^{\circ})^2$ at $z=0.05$ and $\sim (5.3^{\circ})^2$ at $z=0.5$, while for comparison the field of view of a single LOFAR-HBA pointing is $(100^{\circ})^2$. 
The comparison between the high and low amplification model shows that most bright regions inside of halos (considering projected spheres of radius $1.5 R_{\rm vir}$ at all epochs) are identified in both cases (blue regions in the figure), the detectable regions outside of halos strongly depend on the amplification model. In the HA model the LOFAR-HBA survey should detect several bright emission spots within a volume $z \leq 0.5$, in most of cases related to outer accretion regions surrounding galaxy clusters. The best chances of detecting $\sim$ Mpc-sized shocked regions of the WHIM in filaments are for $z \sim 0.1$, which gives a good trade-off between having a good coverage of the necessary short baselines and not being too affected by the cosmological dimming in surface brightness. In the LA model only sparse and more compact knots of bright enough emission are detectable, with a poor reconstruction of the underlying WHIM distribution. At $z=0.5$ (and also up to $z=1$, limited to a very few bright spots) only a magnetic field as high as in our HA model will allow some detection of gas not associated with the virial region of clusters. In the LA case, basically no detection will be possible.

\begin{figure*}
\begin{center}
\includegraphics[width=0.99\textwidth]{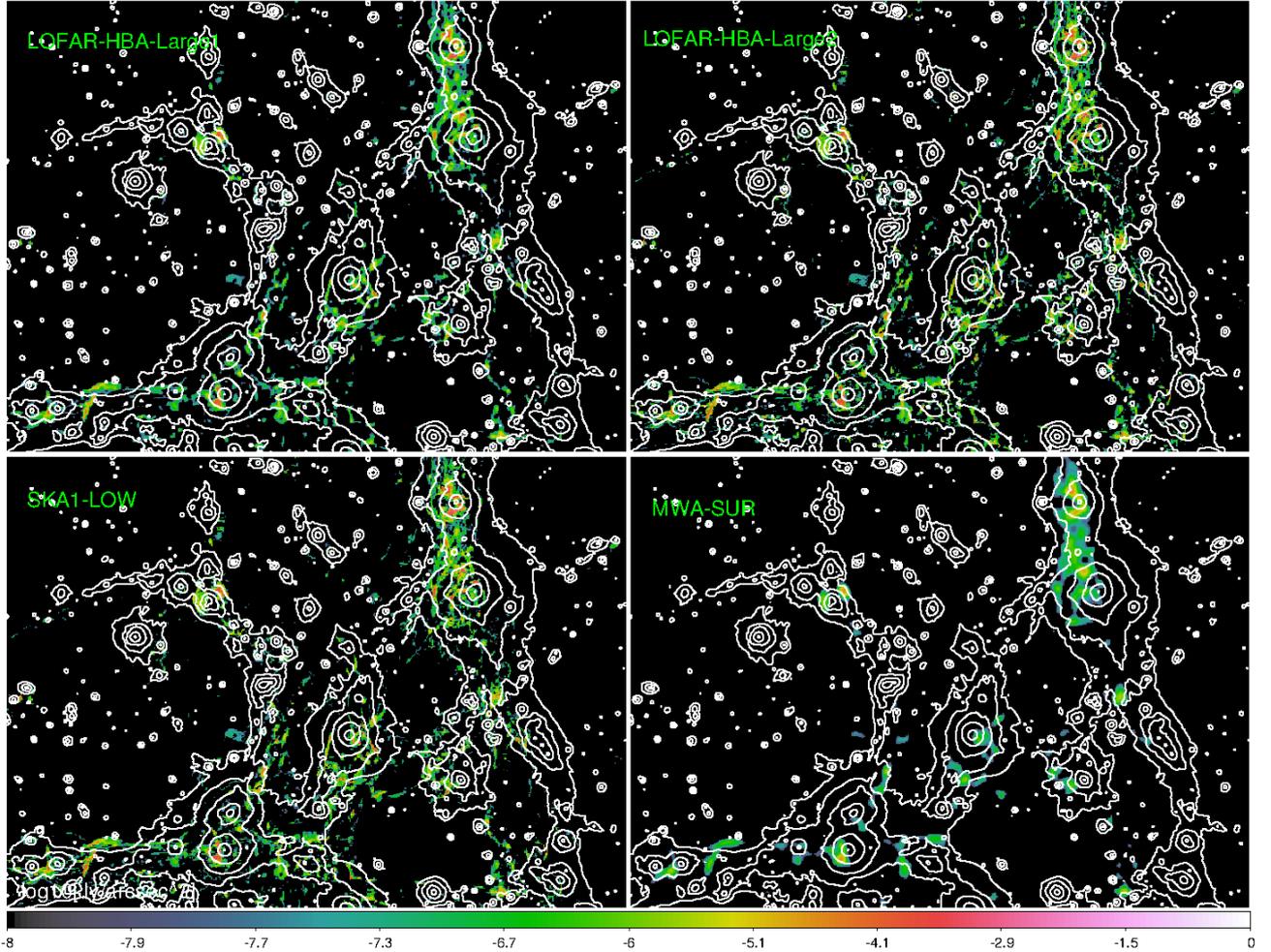}
\caption{Role of resolution in different radio surveys at low frequency. The colours show the detectable emission
by different surveys at low-frequency for the HA model (units of $\log_{\rm 10}$ [Jy/arcsec$^2$]), considering a $5^{\circ} \times 7^{\circ}$ field of view. The emission in each map has been convolved with the corresponding resolution beam. The white contours show the projected gas pressure.}
\label{fig:resol}
\end{center}
\end{figure*}

\begin{figure*}
\begin{center}
\includegraphics[width=0.99\textwidth]{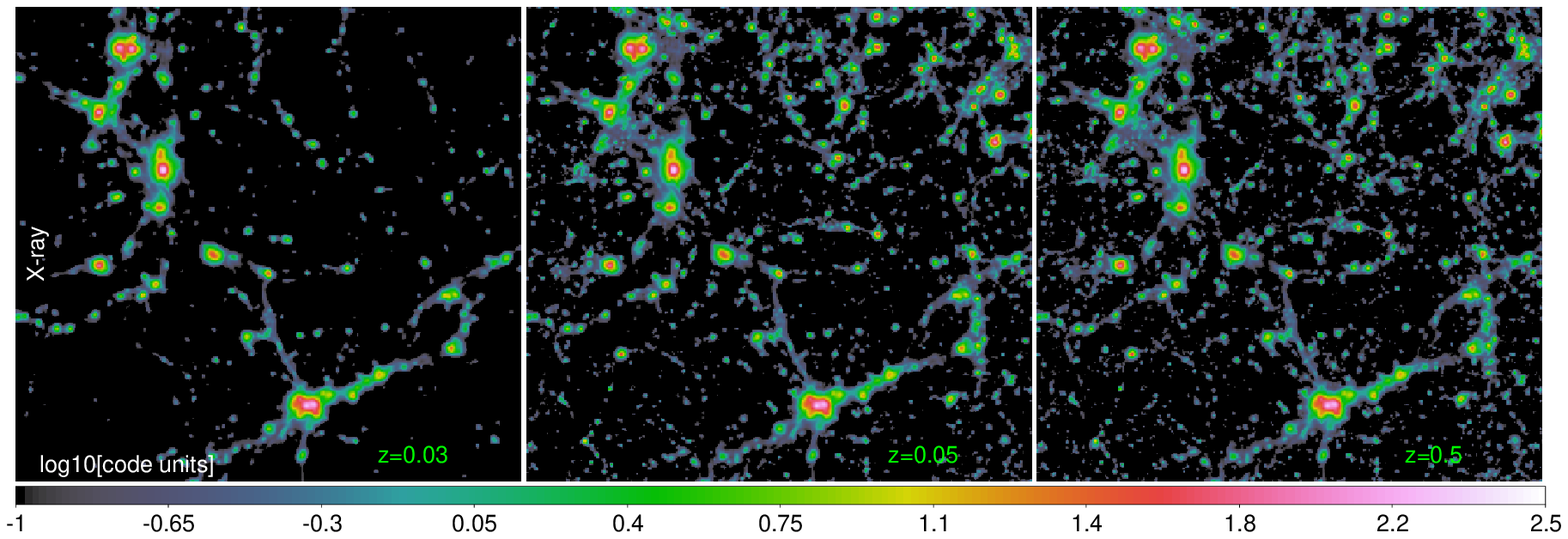}
\includegraphics[width=0.99\textwidth]{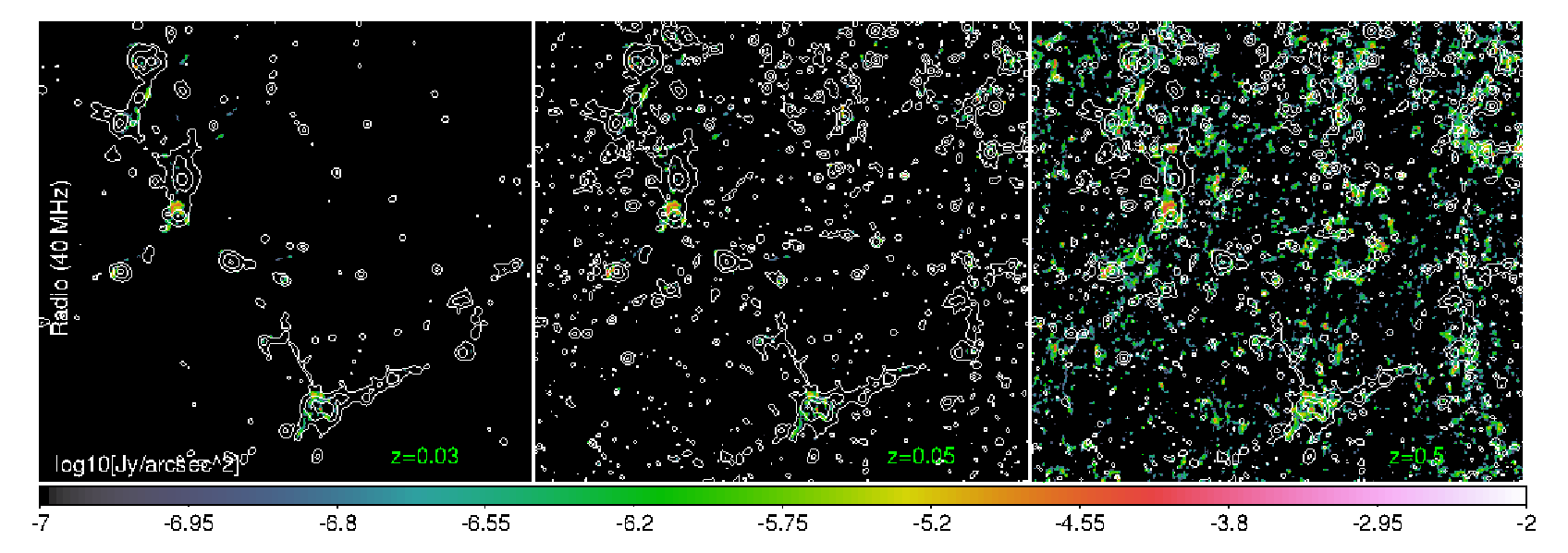}
\caption{Emission from our simulated light cone at $z=0.03$, $z=0.05$ and $z=0.5$, obtained by stacking replicas of our $100 ~\rm Mpc)^3$
and $200 ~\rm Mpc)^3$ simulated volumes. The top panels show the total X-ray emission proxy ($\propto n^2 T^{1/2}$) and the lower panels show the radio emission for LOFAR-LBA (units of $\log_{\rm 10}[\rm Jy/arcsec^2]$) for a projected area of $3.5^{\circ} \times 3.5^{\circ}$, assuming the high-amplification model (HA). The white contours of the X-ray emission proxy are drawn in the lower panels to allow a better  comparison of structures.}
\label{fig:sur_volume}
\end{center}
\end{figure*}


\begin{figure*}
\begin{center}
\includegraphics[width=0.9\textwidth]{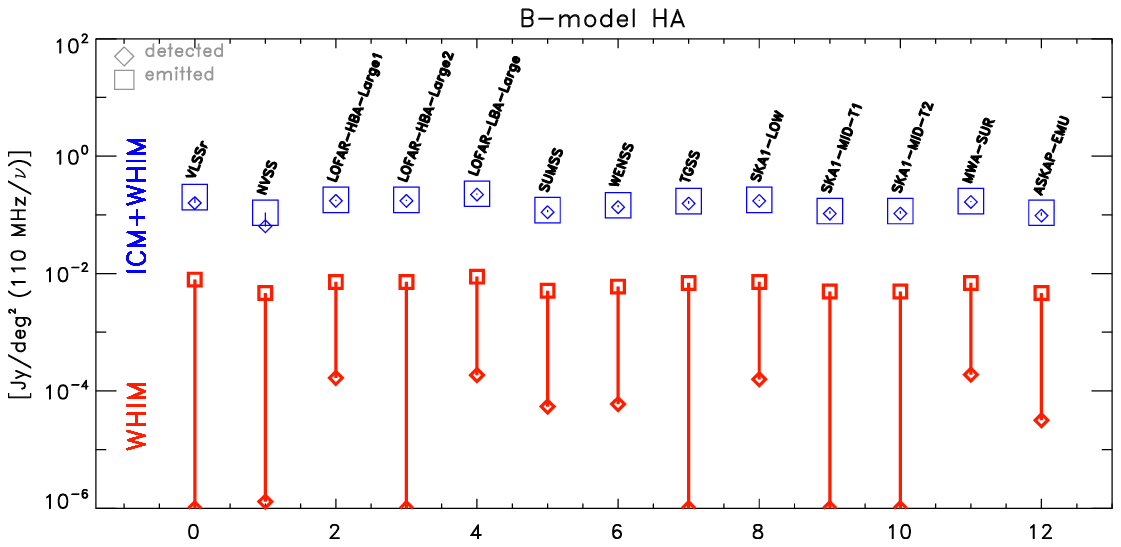}
\includegraphics[width=0.9\textwidth]{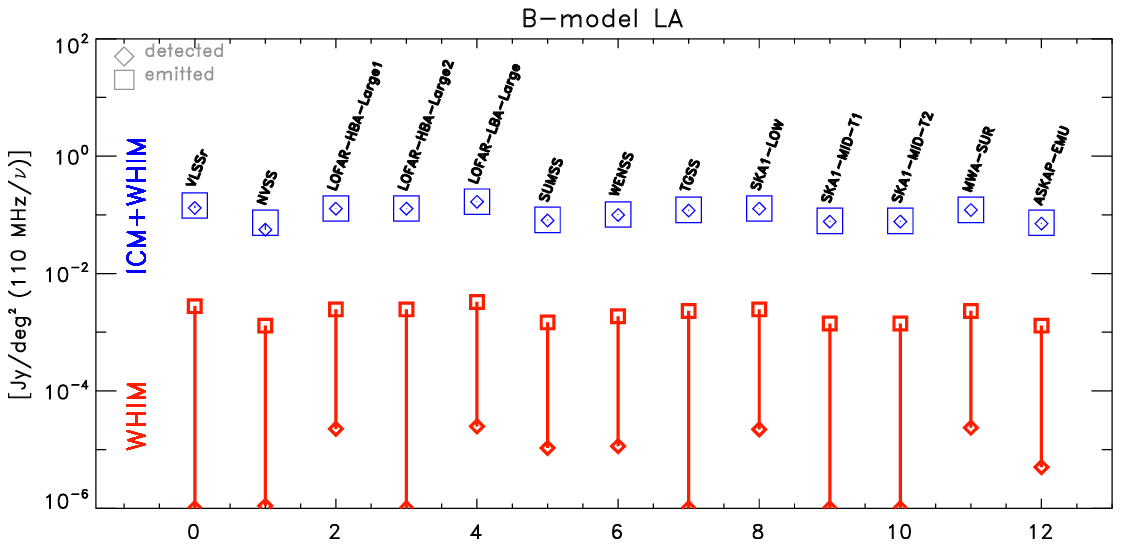}
\caption{Performance of different radio surveys observing a $14^{\circ} \times 14^{\circ}$ region. We simulated all radio emitting volume from $z=0.02$ to $z=0.5$ and present for both models the radio emission from the ICM+WHIM (blue) or from the WHIM only (red) distinguishing the total emission (big squares) and the one detected by each survey configuration (small diamonds). For a better comparison of surveys at different frequencies, all emission have been rescaled by $(110 ~\rm MHz/\nu[MHz])$.} 
\label{fig:sur_stat}
\end{center}
\end{figure*}

\subsection{Performances of radio surveys}
\label{survey}
Next we investigate in depth the chances of detecting the cosmic web by existing and future surveys.
First, we assess the role played by spatial resolution by comparing the maps of four low-frequency surveys, which are the most promising 
for a detection of the cosmic web in a high-amplification scenario (see also discussion below).  Fig.~\ref{fig:resol} shows a $5^{\circ} \times 7^{\circ}$ region 
for the highest resolution volume of $(50 ~\rm Mpc)^3$, observed by LOFAR-HBA at two resolutions, SKA1-LOW and MWA. The spatial resolutions
here span from  $10''$ (SKA1-LOW) to $120''$ (MWA). Each survey should be able to detect a few bright emission regions from the WHIM. However, only having a large enough resolution it is possible to guess the morphology and the spatial orientation of these features, which are otherwise difficult to be distinguished from the possible signal of relic-like emission from distant clusters not yet detected in the X-rays \citep[e.g.][]{2014A&A...565A..13M}.
Furthermore, achieving the highest possible resolution is important for removing point sources and thereby for reducing the confusion noise, which is expected to limit the sensitivity of SKA1-LOW \citep[e.g.][]{2014arXiv1412.6512P}.

Second, we study the performances of all surveys in Tab. 2 using a more realistic setup, where we integrated a long rectangular volume covering  $14^{\circ} \times 14^{\circ}$ in the sky and sampling all cosmic volume from $z=0.02$ to $z=0.5$ (corresponding to a comoving radial distance of $\approx 1.892$ Gpc). 
To do so  we stacked many simulated volumes along the line of sight, starting with a few replicas of the  $(100  ~\rm Mpc)^3$ box to give a better angular resolution at low redshift, and then by stacking several replicas of the  $(200  ~\rm Mpc)^3$ volume to cover all projected volume up to $z=0.5$. 
We first computed the radio emission in the comoving frame of each box at the appropriate redshift, and then ``moved'' each map away from the observer, by applying the cosmological corrections for the surface brightness and the luminosity distance of each redshift, and decrease the pixel size of each box as a function of distance, using a cubic interpolation on the input map. The presence of possible artefacts due to from the periodicity of structures along the line of sight is minimised by shifting each box with random offset in both directions. 
We notice that the first simulated volume along the line of sight is large enough that the change of redshift across the volume cannot be neglected. Therefore, we further subdivided it into 4 slabs (starting at $z \approx 0.023$, $0.029$, $0.035$  and $0.044$), and accordingly computed the cosmological corrections for each of them independently.

Fig.~\ref{fig:sur_volume} shows effect of adding more volume along a projected field-of-view, where we show the proxy of X-ray emission ($\propto n^2 T^{1/2}$) and the radio emission for a subset of the full map created  above, assuming the HA model and for a Large Survey with LOFAR-LBA. Despite the evident appearance of many additional clusters in the field, when our integration region goes from $z=0.03$ to $z=0.5$ the detectable radio emission does not change much, due to the strong cosmological dimming for $z \leq 0.1$. A few brighter spots, correlated with higher redshift merging clusters appear in the full-volume image, yet most of the cosmic web stays below the detectability threshold for LOFAR-LBA (large survey with resolution $25''$). 

To give a more quantitative view of what fraction of the cosmic web can actually be detected by different surveys and for different magnetic field models, we give in Fig.~\ref{fig:sur_stat} the result for the full integrated volume. We show the total emission from all shocked cells, or from the WHIM only after excision of the projected virial volume of each identified cluster or group. In the HA model, the power in the WHIM is $\sim 1$\% of the radio power emitted by clusters, while this is $\sim 0.1$\% in the LA model. Most of the emission in the projected volume comes from clusters and all surveys appear to be able to detect most of it (within a factor $\sim 2$), with the least detection coming from the highest frequency surveys. 
On average, we measure a diffuse radio flux density of $\sim 0.1$ Jy/deg$^2$ (100 MHz/$\nu$) from all shocks in our simulated volume (very similar numbers for both magnetic field model), while  the contribution from the WHIM is $\sim 5 \cdot 10^{-3}-10^{-2}$ Jy/deg$^2$ (100 MHz/$\nu$) in the HA case and $\sim 10^{-3}-5 \cdot 10^{-3}$ Jy/deg$^2$ (100 MHz/$\nu$) in the LA model.

The differences in the performance between surveys becomes more apparent if we look at the residual WHIM emission only. The most satisfactory reconstruction of the WHIM  is obtained with LOFAR (HBA Large 1 and LBA Large surveys), SUMSS, WENSS, SKAl-LOW, SKA1-MID and MWA-SUR, with a fraction of recovered flux which is $1/50-1/100$ of the total WHIM emission. A bigger gap between the total and the detectable emission from the WHIM is measured for the LA case.

\subsection{Filaments connecting clusters}
\label{clusters}

\begin{figure}
\begin{center}
\includegraphics[width=0.498\textwidth]{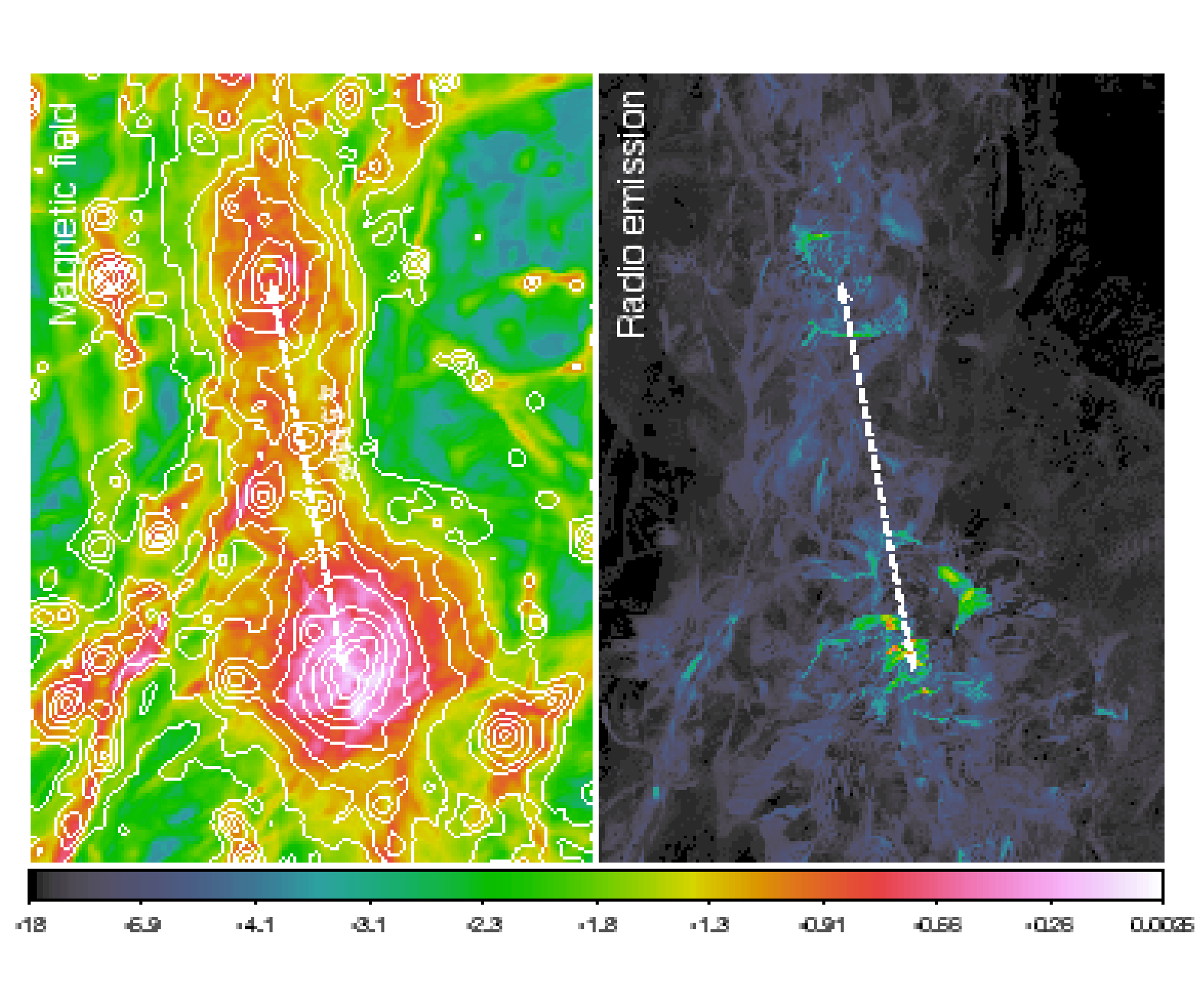}
\includegraphics[width=0.498\textwidth]{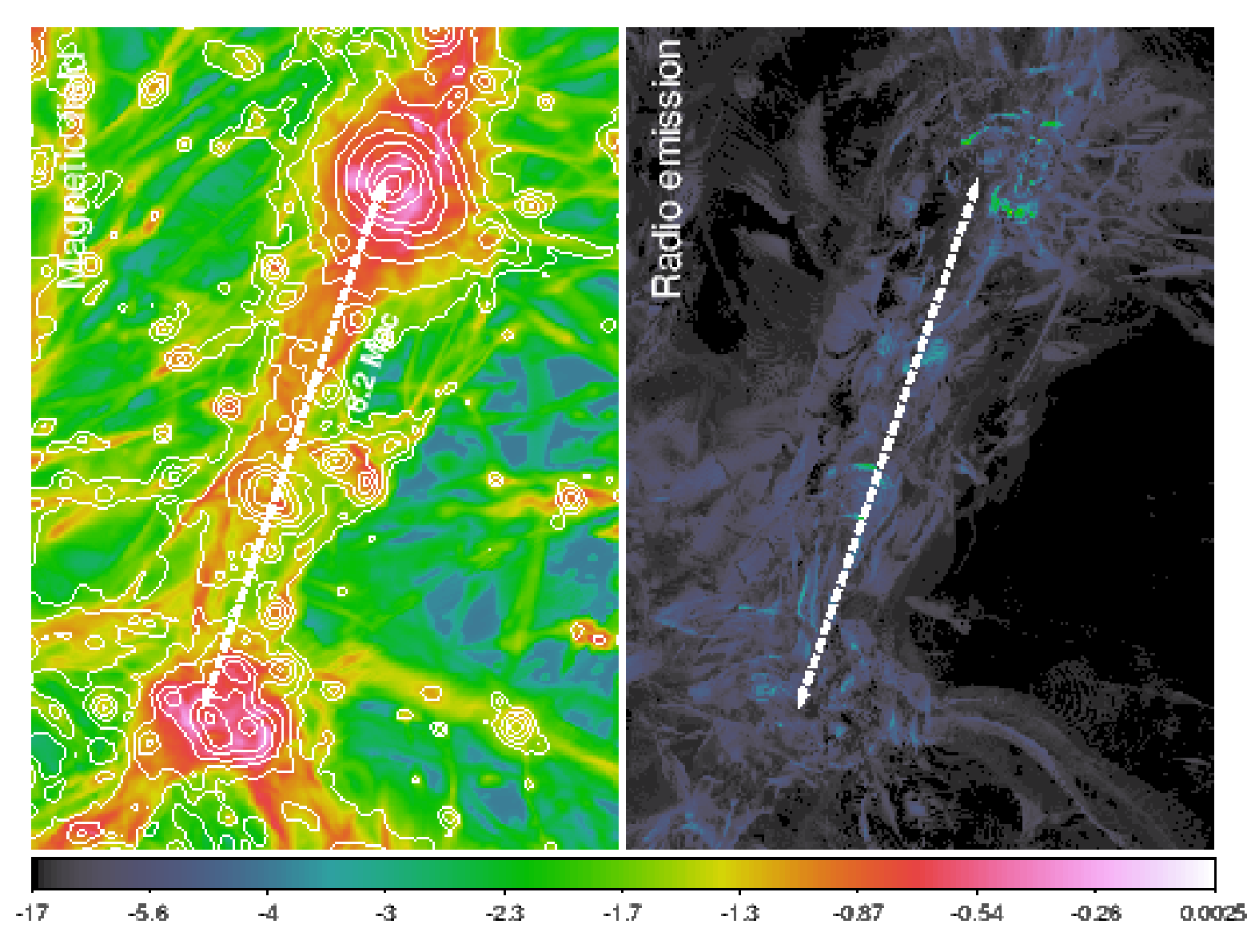}
\caption{Maps of intracluster filaments in the 50 Mpc)$^3$ volume at $z=0.05$. The left panels show the projected mass-weighted magnetic field (HA model, units of $\log_{\rm 10}[\mu G]$) and the white contours give the projected gas density. The right panels show the intrinsic radio emission in the same model ([Jy/arcsec$^2$]).} 
\label{fig:fila_maps}
\end{center}
\end{figure}

\begin{figure}
\begin{center}
\includegraphics[width=0.498\textwidth]{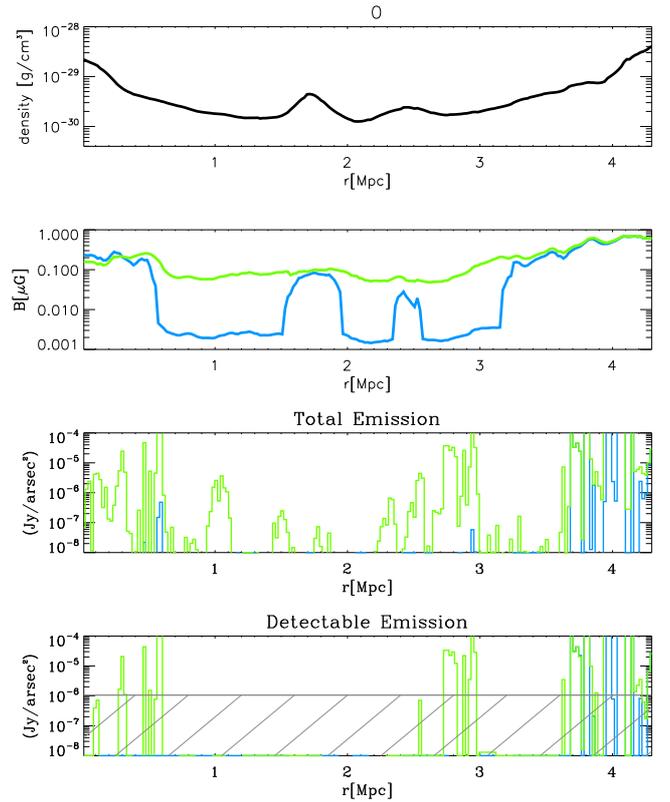}
\caption{Profiles along the white line in the first panel of Fig.~\ref{fig:fila_maps}, showing the trend of mean gas density along the line of sight, magnetic field (each color gives a different model, with blue=LA and green=HA), intrinsic radio emission (same color coding) and detectable radio emission assuming the sensitivity of  LOFAR HBA Large Survey at 120 MHz (resolution 25").}
\label{fig:fila_prof1}
\end{center}
\end{figure}

\begin{figure}
\begin{center}
\includegraphics[width=0.498\textwidth]{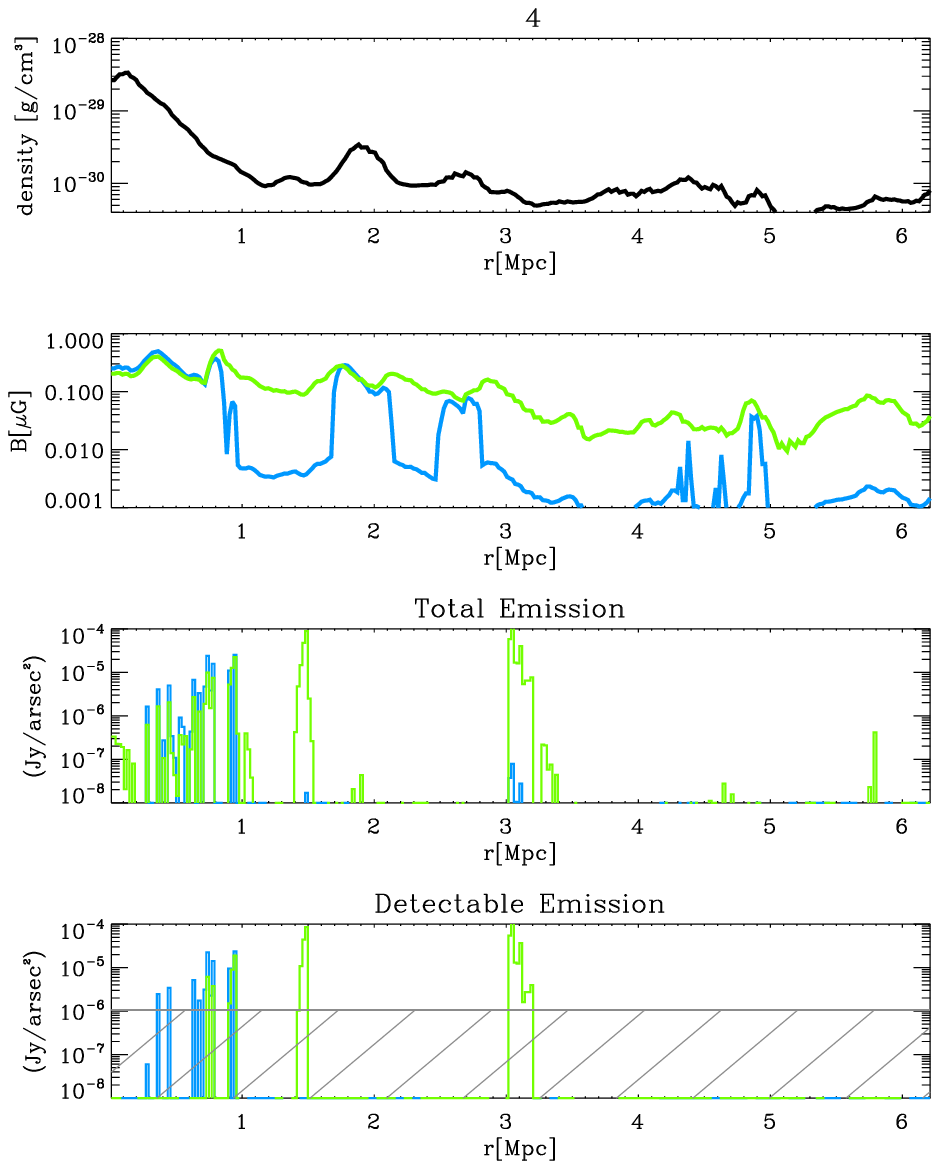}
\caption{Same as Fig.~\ref{fig:fila_prof1}, but for the second cluster association in the Figure.}
\label{fig:fila_prof3}
\end{center}
\end{figure}

Even before the completion of future deep radio surveys, it will be possible to detect the radio emitting WHIM with long exposures targeting galaxy clusters at low frequency. Close pairs of interacting clusters are particularly interesting for this
\citep[e.g.][]{2013A&A...550A.134P,2013ApJ...779..189F}.
Although these objects are much smaller than the $\sim 10-20 ~\rm Mpc$ filaments that can be detected in our simulations, they can give us a first hint of the active mixing between the  ICM and the WHIM. 
First DM-only simulations showed that most of simulated clusters are found to be surrounded by filaments  with typical lengths $\leq$ 20  Mpc/h, and with a weak mass-dependence: clusters with $M \sim 10^{14} M_{\odot}$ are on average surrounded by $\sim 2-3$ filaments while this increases to $\sim 4-5$ for  $M \sim 10^{15} M_{\odot}$ objects \citep[e.g.][]{2005MNRAS.359..272C}. Our recent analysis of the gas properties of the filaments in large cosmological simulations showed that up to $\sim 10^3$ filaments can be identified in $(100  ~\rm Mpc)^3$ at $z=0$, with a decreasing power-law distribution of lengths from one Mpc up to several tens of Mpc (Gheller et al., submitted).  
Deep imaging of the region between close pairs of galaxy clusters is potentially an efficient way to search for cosmic filaments, even if in such close configurations it might be difficult to disentangle the emission from cluster accretion shocks and filaments shocks. \\
In a first exploratory study, we have extracted the position of halos with $\geq 5 \cdot 10^{13} M_{\odot}$ in our most resolved box and have extracted various quantities along the line joining the centres of close clusters.  A more complete survey of the cluster-to-filament connections (involving the cross-correlation of cluster and filament catalogs) will be the subject of future work.  
Here we show a couple of representative cases where some emission from the WHIM can
be detected.  Fig.~\ref{fig:fila_maps} shows the projected maps of magnetic fields (with isodensity contours) and of radio emission at 120 MHz (observed with LOFAR-HBA Large Survey 1) for two pairs of interacting clusters in the  (50 Mpc)$^3$ box. The distances between the centres of the gas halos are $\sim 4.5$ and 6 Mpc, and the connecting bridges are $\sim 1$ Mpc thick. 
The magnetic field along the two filaments is $\sim 0.05-0.1 \mu$G in the HA model and $\sim 0.001 \mu$G in the LA model (except that in a few dense substructures within the filament), as shown by Figs.~\ref{fig:fila_prof1}-\ref{fig:fila_prof3}.
The interaction region of both pairs host significant radio emission, associated with several shocks within the filaments and around the gas substructures internal to it. 
In both cases for a high level of magnetisation a few bright emission regions $\sim 100-200$ kpc long can be detected, while most of the weaker shocks close to the spine of the bridges will stay below the sensitivity of LOFAR. In a low-magnetisation scenario basically no detection from these bridges will be possible, while the relic-like emission inside the main halos remains similar. The presence of radio galaxies can add radio emitting plasma on these bridges, specially if such close pairs are found in a rich cosmic environment.

\subsection{Alternative scenarios}
\label{hsa}
\begin{figure}
\begin{center}
\includegraphics[width=0.495\textwidth]{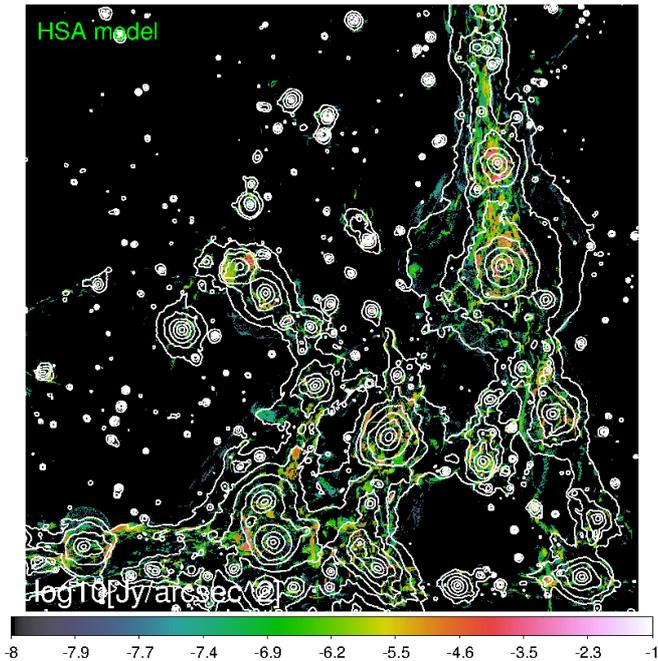}
\caption{Similar to Fig.~\ref{fig:radio_models}, but using the HSA model (high amplification and CR-driven dynamo).}
\label{fig:radio_hsa}
\end{center}
\end{figure}

\begin{figure}
\begin{center}
\includegraphics[width=0.495\textwidth]{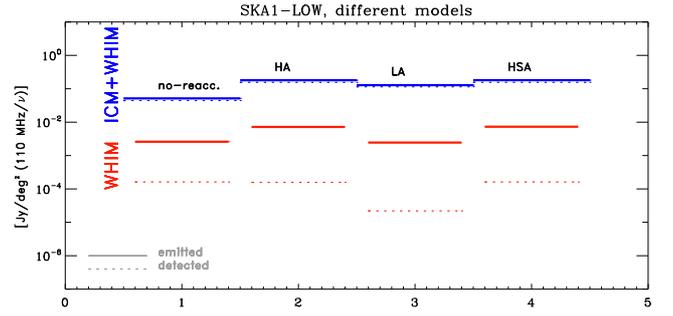}
\caption{Comparison of the emitted and detected radio emission for the same  $14^{\circ} \times 14^{\circ}$ region of Fig.~\ref{fig:sur_stat},  using SKA1-LOW. Each column shows the outcome of different models for the magnetic fields (HA, LA, HSA) and also without the contribution from re-accelerated electrons.}
\label{fig:models}
\end{center}
\end{figure}

We plan to investigate the role played by galaxies and active galactic nuclei in the seeding of large-scale magnetic fields with direct more complex numerical simulations in future work.
Here we briefly discuss a few possible variations of the magnetic field models studied above, which we can easily test with existing runs. 
A possibility suggested in the literature is that cosmic rays accelerated by strong shocks can induce turbulence in the shocks up-stream and trigger substantial magnetic 
field amplification \citep[][]{2012MNRAS.427.2308D, 2013MNRAS.tmp.2295B}. We tested this scenario (labelled HSA) by computing the 
kinetic energy flux which should be dissipated into CRs based on our shock finder and on the acceleration efficiency by \citet[][]{kr13}. This is converted into a magnetic field for each cell, by integrating for the
cell size and the time step of the simulation: $B_{\rm eq,CR}=\sqrt{0.1 E_{\rm CR} 8 \pi}$, i.e. we assume that a $10$\% of the energy of accelerated CR goes into magnetic field amplification. The field obtained in this way is added to the field in the previous HA model.
Fig.~\ref{fig:radio_hsa} shows the radio emission obtained in this scenario: the observed morphologies and power are very similar to the HA case, with the exception of the accretion shocks, where the magnetic field can be amplified by a factor $\sim 10$  and the emission can be increased up to $\sim 10-100$ at accretion shocks around clusters and filaments. However, this extra amplification is often found at densities too low to contribute significantly to the observable radio emission. Fig.~\ref{fig:models} shows the outcome of the HSA model on the emitted and detected radio emission for the same long projected volume analyzed in the previous sections. The total emission in the box, largely dominated by clusters, is the same when the HA and HSA model are compared, and also the total emission from the WHIM remains basically the same.
In any case, for the moment this mechanism remains just a speculation, given the largely unknown acceleration efficiency of CR-protons by cosmological shocks, and the complex details related to the diffusion coefficient of particles in the upstream \citep[see discussions in ][]{2012MNRAS.427.2308D, 2013MNRAS.tmp.2295B}.

Secondly, we tested the case in which the radio emitting electrons are only those freshly accelerated by shocks and there is no contribution from re-accelerated electrons at low Mach number (Sec.~\ref{subsubsec:cr}). Fig.~\ref{fig:models} shows that neglecting shock re-acceleration decreases the total emission by a factor $\sim 2$, by reducing the relic emission in clusters (as already seen in Sec.~\ref{nvss}). Also the WHIM emission is reduced by a similar factor. 

In summary, the presence of additional amplification by cosmic-ray driven turbulence or the presence of pre-existing electrons should affect our estimates only by a very small factor, $\leq 2$ in the total and detectable emission. 

Finally, we notice that the total emitted power from the  WHIM essentially scales linearly with the saturated value of $\xi(M,T)$ for strong shocks (Eq.~\ref{eq:hb}). In this work this is $\xi \sim 0.7-0.9 \cdot 10^{-4}$, of the same order of what usually assumed to explain the observed spectra of supernova remnants \citep[][]{ed11,2014ApJ...789...49F}. Given that the predicted radio spectra of most shocks in the WHIM is flat $P_{\nu} \propto \nu^{-1}$, Eq.~\ref{eq:hb} predicts that for magnetic field lower than $3.2 \mu$G and at low redshift, different combinations of $\xi \cdot B^2$ should give the same emission level.
Considering the level of predicted emission from the WHIM in our runs (Sec.~\ref{survey}), we can therefore reformulate our model prediction in the following way:

\begin{equation}
P_{\rm WHIM}(\nu) \sim 5 \times 10^{-3} \rm {Jy/deg^2}\frac{100 MHz}{\nu} \cdot \frac{\epsilon_{\rm B}}{0.01} \cdot \frac{\xi}{10^{-3}} \ ,
\label{eq1}
\end{equation}
where $\epsilon_{\rm B}$ is the magnetisation level of the WHIM we assumed in the HA model, i.e. $1$\% of the thermal gas energy. This can also be roughly translated into a prescription for the mean magnetic field of the WHIM, assuming a mean density of $n/n_{\rm cr} \approx 10$:

\begin{equation}
P_{\rm WHIM}(\nu) \sim 5 \times 10^{-3} \rm  {Jy/deg^2}\frac{\rm 100 MHz}{\nu}\cdot (\frac{B}{0.05 \mu {\rm G}})^2 \cdot \frac{\xi}{10^{-3}} .
\label{eq2}
\end{equation}
Thus, the level of radio emission we predict for the WHIM can be rescaled for
the simple cases of a different constant magnetisation level of the WHIM  (e.g. assuming a different cosmological seed magnetic field) or  of a revised acceleration efficiency for electrons (e.g. revised acceleration efficiency from supernovae).

\section{Conclusions}

In this work we studied the detectability of the cosmic web using existing and future radio surveys. 
We simulated the acceleration of electrons by cosmological shock waves \citep[][]{hb07} and applied it to large cosmological volumes produced using the MHD version of {\enzo} \citep[][]{va14mhd}. In post-processing we modified the magnetic field as a function of density such as to reproduce the magnetic filed levels observed in galaxy clusters.  We increased the magnetic field according to a high-amplification (HA) or low-amplification (LA) models, which both pushes the magnetic field in clusters close to the observed values (a few $\sim \mu G$) but differ in filaments: in the HA model the magnetic field in filament is set to 1 \% of the thermal gas energy of cells, while in the LA it remains the one from the MHD simulation, i.e. $\leq nG$. 
We produced mock radio observation with the specific parameters of real surveys (including the lack of short baseline, of the spatial resolution and sensitivity of each instrument) and quantified the chance of detecting the WHIM in large projected volumes (up to $z=0.5$). 

Our main results can be summarised as follows:
 \begin{itemize}

 \item Only the high-amplification scenario allows a detection of the tip of the iceberg of the WHIM radio emission, limited to the brightest emission spots in filaments and accretion shocks. The high-magnetisation might come from unresolved small-scale dynamo (or fast growing instabilities in a high plasma $\beta$) and/or from seeding mechanisms not yet included in our MHD runs (e.g. outflows from galaxies, AGN), which push the magnetic field up to $\sim 1$\% of the thermal energy of the WHIM as in clusters (Sec.~\ref{magnetic}). 
 
 \item Low-frequency ($\leq 300$ MHz) observations are best suited to detect the large-scale diffuse emission produced by cosmological shocks, at the scale of cluster accretion shocks or in filaments. These shocks are usually several $\sim$ Mpc wide, are characterised by a flat emission spectrum ($P_{\nu} \propto \nu^{-1 \div -1.5}$) and emit at the level of $\sim  \rm \mu Jy/arcsec^2$ at low redshift, in a high amplification model (Sec.~\ref{survey}).  
 
\item Most of the detectable radio emission is caused by structures at $z \leq 0.1$, because of cosmological dimming. Especially at high frequencies, the detection of too nearby structures of the cosmic web ($z \leq 0.02$) is made difficult by the lack of short baselines.

 \item Of all investigated surveys, the highest fraction of detected flux from the WHIM phase is obtained by LOFAR (HBA-Large survey and LBA-Large), SKA1-LOW and MWA Broadband survey. In all these cases $\sim 10^{-2}$ of the WHIM emission can be detected, mostly associated with shocks internal to filaments, both driven by the accretion of smooth gas in rich environment or driven by supersonic motions internal to filaments. 

 \item The highest resolution accessible to LOFAR ($\leq 25"$) and SKA1-LOW ($\sim 10"$) should enable a much better identification of the morphology of the underlying filamentary environment, compared to MWA, and to exclude the contamination from cluster shocks and active galaxies. High resolution will be also important to reduce the confusion noise (e.g. in the SKA1-LOW).

\item In a low-amplification scenario the detection chance of the above surveys will be reduced by $\sim 10$, and only a very limited number of bright spots (mostly associated with dense substructures within filaments) will be detectable, offering a poor tracing of the WHIM distribution.

 \item We estimate a background of unresolved radio emission from the WHIM in the range $\sim 10^{-3}-10^{-2} \rm Jy/deg^2 (100 MHz/\nu)$, depending on the amplification model (Eq.~\ref{eq1}-\ref{eq2}). 
 
 \item Clusters interacting on short distances ($\sim 4-6$ Mpc) can be connected by magnetised bridges, hosting internal shocks that can be observed if the magnetisation level is $\sim 1$\% of the thermal gas (Sec.~\ref{clusters}).
 
 \item Our modeling does not yet cover the impact of magnetised outflows from galaxies and active galactic nuclei in a proper way. We investigated a few alternative scenarios for the amplification of magnetic field and for the acceleration of particles at strong shocks, finding only minor differences with respect to the main results of our baseline model (Sec.~\ref{hsa}). 

\item Radio emission from secondary electrons injected by hadronic collisions is negligible in the WHIM (Appendix).

\end{itemize}

An important caveat in our analysis is that we assume that DSA can operate at all density and magnetisation levels in our simulated volume, which may not be true for very low levels of magnetic fields. Recent results from Particle In Cell simulations (PIC) pointed out that more important for the onset of DSA at non-relativistic shock is the topology of the upstream magnetic field, rather than its initial value. In particular, if the upstream magnetic field is aligned with the shock normal (i.e. parallel shocks), the rapid development of the Weibel instability might amplify the magnetic field and enables the formation of a shock precursor leading to DSA \citep[][]{2012JCAP...07..038C,2014arXiv1401.7679C,guo14}. For this reason, the presence of shock-accelerated electrons in the WHIM  is a realistic possibility.

An important aspect of the above scenario is that both a detection or a non-detection of the cosmic web in radio within the next decade will be helpful to put constraints on the magnetic field in the WHIM, given that to a  first approximation the radio emission from the WHIM should scale as $P_{\rm \nu} \propto \xi B^{2}$ (Eq.\ref{eq1}-Eq.\ref{eq2}). The lack of detection in radio from large observed volume will limit the magnetic energy in the WHIM to values much below $\sim 1$ \% of the thermal gas energy in these environment. In this case, the effect of cosmic magnetic fields on the propagation of UHECRs is predicted to be negligible \citep[][]{2003PhRvD..68d3002S,2005JCAP...01..009D}, and so is the effect of magnetic fields in producing axion-like particle oscillations \citep[][]{2012PhRvD..86g5024H}.

On the other hand, in the case of systematic detection of emission patches connected to the WHIM, the analysis of their morphology and large-scale distribution will be important to  distinguish competing scenarios. For example the detection of coherent magnetic fields 
on $\gg \rm Mpc$ scales will favour a distributed amplification process as an obliquitous small-scale dynamo.  Our models show that if the growth of cosmic web is linked to the thermal gas energy of the WHIM, the cosmic web should be maximally emitting in radio at moderately low redshift ($z \leq 0.1$). Any unambiguous detection of radio emission from much higher redshift will further discriminate between seeding models and favour a strong magnetisation activity of large-scale structures by high-$z$ star-forming galaxies.  
 
Beside continuum radio emission, other complementary techniques should be able to probe the existence of the cosmic web, using different
effects: polarised radio emission \citep[e.g.][]{2015arXiv150100389G}, Faraday Rotation \citep[e.g.][]{2014ApJ...790..123A,2015arXiv150100321B,2015arXiv150102298T}, neutral hydrogen emission \citep[e.g.][]{2015arXiv150101077P},  absorption and emission features of highly ionised elements in the WHIM \citep[e.g.][]{2012arXiv1207.2745B} and fast radio bursts \citep[][]{2014ApJ...797...71Z,2014arXiv1412.4829D}.  Considering the extreme difficulty of getting the expected weak signal from the cosmic web, the use of statistical and stacking techniques will increase the chances of success \citep[e.g.][]{2015arXiv150100415V,2015arXiv150100390S}.
%

\section{Acknowledgments}

Computations described in this work were performed using the {\enzo} code (http://enzo-project.org), which is the product of a collaborative effort of scientists at many universities and national laboratories. We gratefully acknowledge the {\enzo} development group for providing extremely helpful and well-maintained on-line documentation and tutorials.\\
We acknowledge the usage of computing time at CSCS, for the allotted CHRONOS project in 2014, and the precious support by  CSCS-ETH in Lugano (www.cscs.ch). We also 
acknowledge PRACE for awarding us access to CURIE-Genci based in France at Bruyeres-le-Chatel in 2013. The support of the TGC Hotline from the Centre CEA-DAM Ile de France to the technical work is gratefully acknowledged.
F.V. and M. B. acknowledge the  usage of computational resources on the JUROPA cluster at the at the Juelich Supercomputing Centre (JSC), under project no. 5018, 5984, 5056 and 7006. F.V. and M.B. acknowledge support from the grant FOR1254 from the Deutsche Forschungsgemeinschaft. 
We thankfully acknowledge R. Pizzo and H. R\"{o}ttgering of fruitful scientific discussion, and I. Prandoni for carefully reading the manuscript.
We acknowledge the use of the online "Cosmology Calculator" by \citet{2006PASP..118.1711W}.

\bibliographystyle{aa}
\bibliography{franco} 
1200

\appendix

\section{The contribution from secondary electrons}

We briefly report here on the contribution from secondary electrons injected by hadronic proton-proton collision \citep[][]{bl99,de00} to the total radio emission. The injection rate of secondary electrons is computed from the pool of CR-protons simulated at run-time by our {\enzo} cosmological simulation using PPM and a two-fluid model for modeling cosmic rays, as in \citet{va14curie}. For this we use the formalism by \citet{de00}, assuming a constant energy spectrum of $\alpha=1.1$ ($N(E) \propto E^{-\alpha}$) for cosmic ray protons. The final budget of cosmic rays in the simulation follows from the assumed acceleration efficiency at shocks, which we took from \citet[][]{kr13} and renormalize downwards of a factor 10 to be consistent with the upper limits from the \emph{Fermi} satellite \citep[][]{fermi14}. 
Here we use the result from a (300 Mpc)$^3$ volume simulated with $2048^3$ cells and DM-particles. This run makes does not use MHD and therefore we have to assume the magnetic field entirely in post-processing. Similar to the main articles, we compute the gas energy in all cells and compute the magnetic field in a HA and LA model. 

\begin{figure}
\begin{center}
\includegraphics[width=0.495\textwidth]{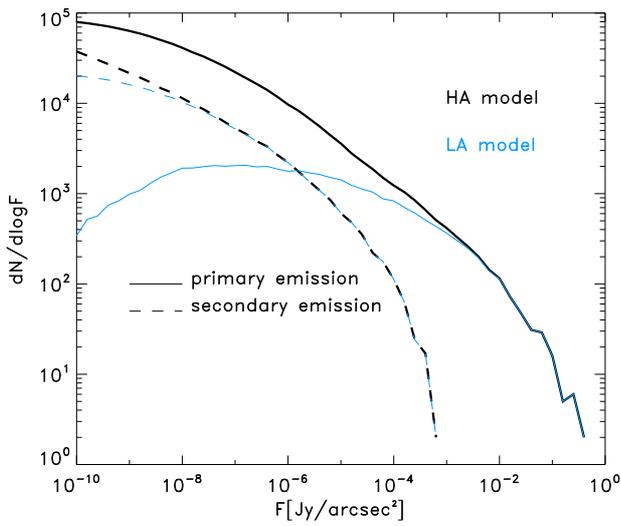}
\caption{Distribution of radio emission from the HA and the LA model in a cosmological run with CR-protons and including the effect of ``secondary'' radio emission from electrons injected by hadronic collision.}
\label{fig:secondary}
\end{center}
\end{figure}

\begin{figure*}
\begin{center}
\includegraphics[width=0.95\textwidth]{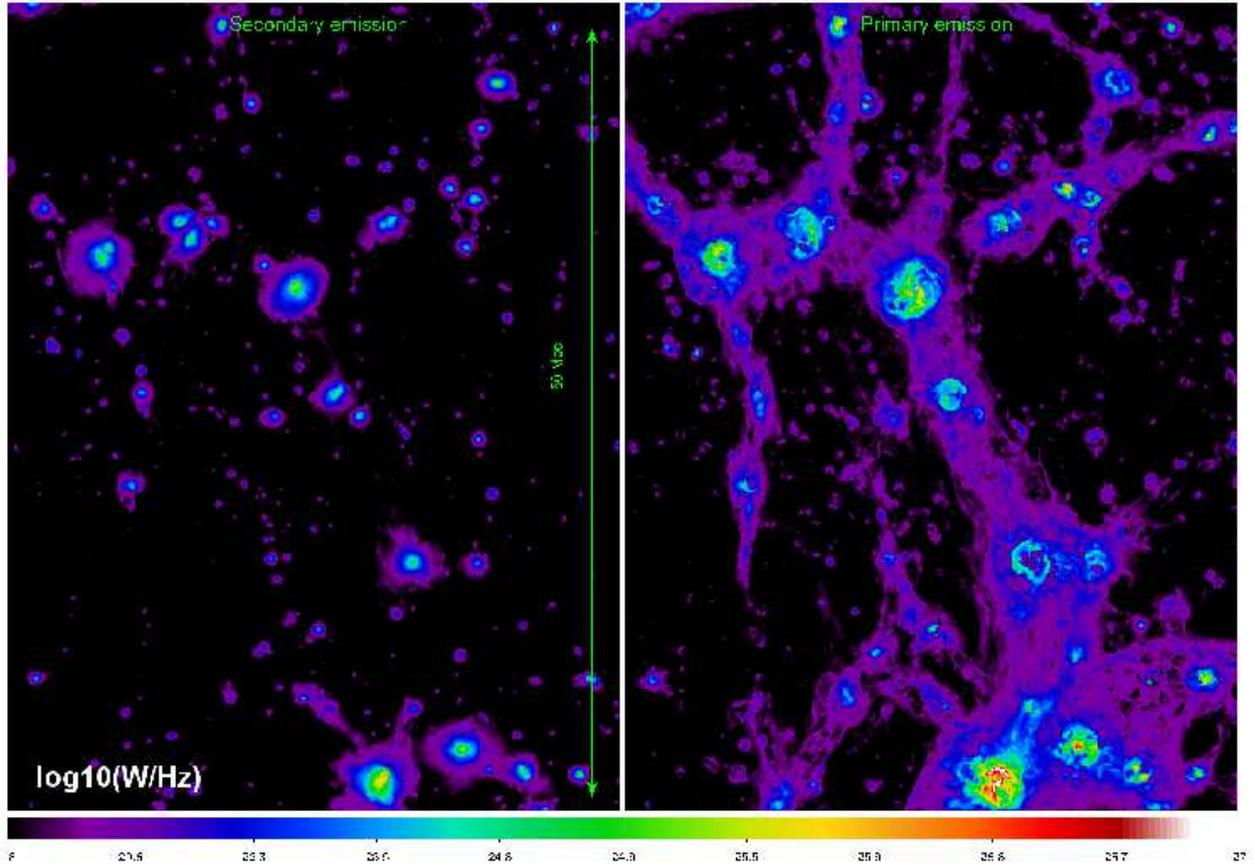}
\caption{Total radio emission from primary electrons injected by shocks (right) and from secondary electrons injected by hadronic collisions (left)  in a subvolume of our $2048^3$ simulation with CR-physics \citep[][]{va14curie}. The color bar gives the emission in  in units of $[log_{\rm 10} \rm W/Hz]$.}
\label{fig:secondary2}
\end{center}
\end{figure*}

Fig.~\ref{fig:secondary} shows the outcome of these two acceleration mechanisms for the HA model: the contribution from secondary electrons to the diffuse cosmic web  is everywhere negligible, i.e. only a few percent of the primary contribution. The emission gets stronger in the centre of the clusters owing to the increase with gas density, and only forms rather low-power radio-halos \citep[e.g.][]{de00}. Outside the central cluster regions, however, this signal usually gets much smaller than the primary emission from strong accretion shocks.
The distribution function of pixels in the maps (Fig.~\ref{fig:secondary2}) shows that
by far the primary contribution dominates the high-end of the emission tail in both magnetic field model. However, the secondary contribution becomes the dominant one in the low-brightness end of the emission distribution in a LA scenario, due to drop of primary emission from the WHIM outside of halos and to the fact that the secondary emission is nearly unaltered in the LA scenario since it mostly comes from well within halos. The confusion caused by the secondary emission from galaxy clusters should become more important 
at high redshift, where massive bright clusters can dominate the low power background emission compared to the filaments, in the LA scenario. 
Overall, we conclude that regardless of the assumed amplification model, the brightest emission from the cosmic web should be originated by primary electrons accelerated by  cosmic shocks, and that the signal from secondary particles can be relevant only at very low values of surface brightness {and at high redshift}, in a low magnetisation scenario of the WHIM. 

\end{document}